\documentclass[10pt,conference]{IEEEtran}

\usepackage{amsmath,amsfonts}

\usepackage{hyperref}
\usepackage{subcaption}
\newsavebox{\mybox}

\usepackage{amsthm}
\usepackage{mdframed}
\usepackage{centernot}
\usepackage{comment}
\usepackage[table]{xcolor}

\newcommand{\Dd}{\mathcal{D}}

\newcommand\vd[2]{d_{i, p}}
\newcommand{\set}[1]{\left\{ #1 \right\}}

\newcommand{\toolname}{\textsc{HyFair}\xspace}
\usepackage{tikz}

\newtheorem{definition}{Definition}[section]

\definecolor{gold}{rgb}{0.99,0.78,0.07}
\usetikzlibrary{arrows,shapes,snakes,automata,backgrounds,positioning,decorations.pathmorphing,
	decorations.markings,calc}

\tikzstyle{dtreenode}=[draw=blue!10!gray,rounded rectangle, minimum size=5mm,fill=blue!10!white]
\tikzstyle{dtreeleaf}=[draw=black!60,minimum width=1cm,minimum height=0.4cm,rectangle,fill=blue!50!white]
\tikzset{every loop/.style={looseness=7}}
\tikzset{
	gluon/.style={decorate,draw=black,
		decoration={coil,amplitude=1pt, segment length=5pt}}
}
\tikzset{
	gluon1/.style={decorate,draw=black,
		decoration={coil,amplitude=3pt, segment length=3pt}}
}
\tikzset{
	gluonew/.style={decorate,draw=black,
		decoration={coil,amplitude=1pt, segment length=2pt}}
}

\tikzset{bicolor/.style args={#1 and #2 and #3}{
		path picture={
			\tikzset{rounded corners=0}
			\fill [#1] (path picture bounding box.south west)
			rectangle
			($(path picture  bounding box.north west)!#3!(path picture bounding
			box.north east)$);
			\fill [#2]
			($(path picture bounding box.south west)!#3!(path picture bounding
			box.south east)$)
			rectangle (path picture bounding box.north east);
}}}

\tikzset{tricolor/.style args={#1 and #2 and #3 and #4 and #5}{
		path picture={
			\tikzset{rounded corners=0}
			\fill [#1] (path picture bounding box.south west)
			rectangle
			($(path picture  bounding box.north west)!#4!(path picture bounding
			box.north east)$);
			\fill [#2]
			($(path picture bounding box.south west)!#4!(path picture bounding
			box.south east)$)
			rectangle
			($(path picture  bounding box.north west)!#5!(path picture bounding
			box.north east)$);
			\fill [#3]
			($(path picture bounding box.south west)!#5!(path picture bounding
			box.south east)$)
			rectangle (path picture bounding box.north east);
}}}

\usepackage[many]{tcolorbox}
\tcbuselibrary{listings,skins}

\lstdefinestyle{mystyle}{
  xleftmargin=0pt,
   basicstyle={\footnotesize\ttfamily},
   aboveskip=3mm,
   belowskip=3mm,
   keywordstyle=\bfseries,
   showstringspaces=false,
  escapechar=?,
  language=Java
}
\definecolor{code_indent}{HTML}{CCCCCC}

\newtcblisting{mylisting}[2][]{
  arc=0pt, outer arc=0pt,
  listing only,
  listing style=mystyle,
  title={\large #2},
  #1
}

 \definecolor{dkgreen}{rgb}{0,0.6,0}
 \definecolor{gray}{rgb}{0.5,0.5,0.5}
 \definecolor{mauve}{rgb}{0.58,0,0.82}

\definecolor{cadmiumgreen}{rgb}{0.0, 0.42, 0.24}
\definecolor{verde}{rgb}{0.25,0.5,0.35}
\definecolor{jpurple}{rgb}{0.5,0,0.35}
\definecolor{darkgreen}{rgb}{0.0, 0.2, 0.13}

\usepackage[ruled,vlined,linesnumbered,lined]{algorithm2e}
\usepackage{algpseudocode}

\usepackage{mathtools,xparse}

\usepackage{tabu}
\usepackage{multirow}

\usepackage{pifont}
\usepackage{enumerate}
\usepackage[shortlabels]{enumitem}

\usepackage{pgfplotstable}
\usepackage{pgfplots}

\usepackage[noframe]{showframe}
\usepackage{framed}

 {\endMakeFramed}
 \definecolor{shadecolor}{gray}{0.85}

\definecolor{bgblue}{RGB}{245,243,253}
\definecolor{ttblue}{RGB}{91,194,224}

\mdfdefinestyle{mystyle}{%
  rightline=true,
  innerleftmargin=10,
  innerrightmargin=10,
  outerlinewidth=3pt,
  topline=false,
  rightline=false,
  bottomline=false,
  skipabove=\topsep,
  skipbelow=false
}

\newtcolorbox{myboxi}[1][]{
  breakable,
  title=#1,
  colback=white,
  colbacktitle=white,
  coltitle=black,
  fonttitle=\bfseries,
  bottomrule=0pt,
  toprule=0pt,
  leftrule=3pt,
  rightrule=3pt,
  titlerule=0pt,
  arc=0pt,
  outer arc=0pt,
  colframe=black!50,
}

\newtcolorbox{myboxii}[1][style=mystyle]{
  breakable,
  freelance,
  colback=white,
  colbacktitle=white,
  coltitle=black,
  fonttitle=\bfseries,
  bottomrule=0pt,
  boxrule=0pt,
  colframe=white,
  after skip=0pt,
  overlay unbroken and first={
    \draw[white!75!black,line width=3pt]
    ([yshift=-9pt]frame.north west) --
    ([yshift=9pt]frame.south west);
  },
}

\usepackage{wrapfig}

\begin{document}

\title{Uncovering Discrimination Clusters: Quantifying and Explaining Systematic Fairness Violations}
\author{
\IEEEauthorblockN{Ranit Debnath Akash}
\IEEEauthorblockA{
University of Illinois Chicago, USA\\
rakas@uic.edu
}
\\
\IEEEauthorblockN{Ashutosh Trivedi}
\IEEEauthorblockA{University of Colorado Boulder, USA\\
ashutosh.trivedi@colorado.edu
}
\and
\IEEEauthorblockN{Ashish Kumar}
\IEEEauthorblockA{Pennsylvania State University, USA\\
azk640@psu.edu}
\\
\IEEEauthorblockN{Gang Tan}
\IEEEauthorblockA{Pennsylvania State University, USA\\
gtan@psu.edu}
\and
\IEEEauthorblockN{Verya Monjezi}
\IEEEauthorblockA{University of Illinois Chicago, USA\\
vmonj@uic.edu
}
\\
\IEEEauthorblockN{Saeid Tizpaz-Niari}
\IEEEauthorblockA{
University of Illinois Chicago, USA\\
saeid@uic.edu}
}

\IEEEtitleabstractindextext{
\begin{abstract}
Fairness in algorithmic decision-making is often framed in terms of \emph{individual fairness}, which requires that similar individuals receive similar outcomes. A system violates individual fairness if there exists a pair of inputs differing only in protected attributes (such as race or gender) that lead to significantly different outcomes---for example, one favorable and the other unfavorable. While this notion highlights isolated instances of unfairness, it fails to capture broader patterns of \emph{clustered discrimination} that may affect entire subgroups. 

We introduce and motivate the concept of \emph{discrimination clustering}, a generalization of individual fairness violations. Rather than detecting single counterfactual disparities, we seek to uncover regions of the input space where small perturbations in protected features lead to \emph{$k$-significantly distinct clusters} of outcomes. That is, for a given input, we identify a local neighborhood---differing only in protected attributes---whose members’ outputs separate into many distinct clusters. These clusters reveal significant arbitrariness in treatment solely based on protected attributes, exposing patterns of algorithmic bias that elude pairwise fairness checks. 

We present \toolname, a hybrid technique that combines formal symbolic analysis (via SMT and MILP solvers) to certify individual fairness with randomized search to discover discriminatory clusters. This combination enables both formal guarantees---when no counterexamples exist---and the detection of severe violations that are computationally challenging for symbolic methods alone. Given a set of inputs exhibiting high $k$-discrimination, we further introduce a novel explanation method that generates interpretable, decision-tree-style artifacts. 

Our experiments show that \toolname outperforms state-of-the-art fairness verification and local explanation methods. It reveals that some benchmarks exhibit substantial discrimination clustering, while others show limited or no disparities with respect to protected attributes. It also provides intuitive explanations that support understanding and mitigation of unfairness.
\end{abstract}
}
\maketitle

\IEEEdisplaynontitleabstractindextext

\section{Introduction}
\label{sec:intro}

The availability of big data, performant training algorithms, and specialized hardware has made deep feed-forward neural networks (DNNs)~\cite{GoodBengCour16} a foundational component of modern software systems. DNNs are now routinely deployed in socio-economic decision-making tasks, including risk assessment for criminal reoffense~\cite{berk2014forecasts}, hiring and recruitment~\cite{AIinHR}, income prediction for loans~\cite{AIinLoan}, and facial recognition~\cite{buolamwini2018gender}. 
However, these models are often opaque and highly non-linear, making them prone to unjustified disparities---cases where inputs differing only in protected attributes (e.g., race, gender) yield significantly different outputs. Such disparities raise serious concerns about fairness, particularly in high-stakes domains where automated systems may disproportionately allocate opportunities or resources across social groups.

Traditional fairness verification methods~\cite{10.1109/ICSE48619.2023.00134,khedr2023certifair} 
attempt to certify the absence of unfairness through exhaustive search, while fairness testing approaches
~\cite{agarwal2018automated,udeshi2018automated,zhang2020white,9793943,10.1145/3468264.3468537,10.1145/3540250.3549093,peng2022fairmask} 
rely on randomized exploration to uncover counterexamples. However, verification can be computationally prohibitive for rich fairness specifications, and testing often struggles in regions of the input space with low gradient signals or fairness plateaus. Moreover, most existing work focuses on isolated violations of individual fairness and fails to capture broader patterns of systematic bias.

In this work, we propose a new framework for \emph{discrimination clustering}, which generalizes individual fairness violations by uncovering k-significantly distinct clusters of outcomes within counterfactual neighborhoods---regions of the input space differing only in protected attributes. Our hybrid approach, \toolname, combines formal symbolic analysis with randomized search to both certify fairness and quantify the strength of unfairness. By doing so, we aim not just to detect fairness violations but to quantify and explain the underlying systematic disparities that may otherwise remain hidden.

\vspace{0.25em}\noindent \textbf{Modeling Fairness Violations.}
When analyzing decision-support software through the lens of fairness, it is common to model such systems as binary classifiers, where outputs are categorized as either \emph{favorable} or \emph{unfavorable}. 
The features of inputs under consideration can be partitioned into \emph{protected attributes} (e.g., race, gender identity, disability status) and \emph{non-protected attributes} (e.g., income, work experience, education level). 
Under a standard formulation of equality of opportunity~\cite{barocas-hardt-narayanan}, fairness requires that decision outcomes depend only on relevant, non-protected features so that similar individuals, differing only in protected attributes, receive similar outcomes.
This notion underlies the search for individual discrimination (ID), where an individual and their counterfactual---differing only in a protected attribute---receive different outcomes. 
Such individual fairness violations, also known as discriminatory instances, have been widely studied~\cite{agarwal2018automated,udeshi2018automated,zhang2020white,9793943,10.1145/3468264.3468537}, particularly in the software testing community. 
However, this binary framing overlooks complex patterns of unfairness, such as when neighborhoods of inputs, differing only in protected attributes, lead to multiple divergent and arbitrary outcomes.

To capture these richer fairness violations, we propose a quantitative generalization of counterfactual fairness.
We define a system to be $k$-discriminant if, within a group of $K$ counterfactual inputs---records that differ only in protected attributes---the system produces $2 \leq k \leq K$ distinct outcomes. Much like how k-means clustering partitions data based on feature similarity, this formulation identifies clusters of discriminatory behavior: localized regions in the input space where variations in protected features alone lead to outcome groupings that are meaningfully separated. We refer to this phenomenon as \emph{discrimination clustering}. This notion offers multiple advantages over prevalent individual discrimination. While exiting tools can generate hundreds of thousands of IDs, $k$-discrimination allows us to prioritize test cases based on the severity of discrimination.
Finding $k$-discriminants also helps uncover worst-case scenarios where a DNN makes highly inconsistent (or arbitrary) decisions for similar individuals based on their protected attributes.
These instances often show a structured pattern of systematic disparities~\cite{creel2022algorithmic} that moves beyond isolated violations of individual fairness.

\vspace{0.25em}\noindent \textbf{\toolname: Characterizing $k$-Discriminant Clusters.}
Given the pervasiveness of their socio-economic-critical applicability, fairness testing and verification for DNNs have received considerable attention~\cite{dwork2012fairness,albarghouthi2017fairsquare,kusner2017counterfactual,Monjezi2023InformationTheoreticTA}. Verification approaches~\cite{10.1109/ICSE48619.2023.00134,khedr2023certifair,DBLP:conf/cav/LiWW23} aim to provide a mathematical proof of fairness through exhaustive exploration, while testing-based approaches~\cite{agarwal2018automated,udeshi2018automated,zhang2020white,9793943,10.1145/3468264.3468537,10.1145/3540250.3549093,peng2022fairmask} seek to increase trust by randomized discovery of discrimination-related bugs.
Exhaustive formal search excels at proving the absence of violations but struggles to scale for complex discrimination properties, and may generate large volumes of counterexamples, overwhelming human analysts. In contrast, randomized search scales better but often performs poorly when exploring flat or low-signal regions of the outcome space.

Both strategies offer complementary strengths. Randomized approaches scale well and have been effective in uncovering $2$-discriminant instances (individual fairness violations), but they cannot guarantee the absence of discrimination. As Dijkstra might put it, testing can show the presence of discrimination, but never its absence.
Meanwhile, the formal methods community~\cite{10.1109/ICSE48619.2023.00134,khedr2023certifair,DBLP:conf/cav/WangLW22,DBLP:conf/cav/LiWW23} has advanced verification approaches using constraint solvers (e.g., abstract interpretation, SMT) to exhaustively prove the absence of individual discrimination, i.e., the lack of $2$-discriminant counterfactuals.
However, such approaches face scalability challenges when generalized to quantitative fairness, such as $k$-discriminant clusters.

\textit{
This paper presents \toolname, a hybrid approach for the fairness analysis of DNNs that combines formal verification with randomized exploration to uncover, quantify, explain, and mitigate \emph{clustered patterns of discrimination}. The goal is to support developers of data-driven software in diagnosing and addressing fairness violations that arise from significant arbitrariness in the behavior of DNNs.}

Our hybrid approach first uses MILP solvers~\cite{10.1109/ICSE48619.2023.00134,Fischetti2018} to either certify fairness or find counterexamples witnessing $2$-discriminant instances. Our results show that \toolname significantly outperforms \textsc{Fairfy}~\cite{10.1109/ICSE48619.2023.00134}, the state-of-the-art technique in terms of finding more individual discrimination instances quicker (RQ1).
It then performs a local randomized search around these seed counterexamples to identify maximum $k$-discriminant clusters, i.e., structured regions in the input space where protected attribute perturbations lead to multiple distinct outcomes.
We use a simulated annealing search to find inputs with the maximum $k$-discrimination that outperforms baseline randomized search strategies (RQ2).

\vspace{0.25em}\noindent
\textbf{\toolname: Debugging $k$-Discriminant Clusters.}
We next aim to understand the common conditions underlying the set of edge-case inputs \{$\mathbf{a}_1, \ldots, \mathbf{a}_k$\} that witness a $k$-discriminant cluster, i.e., a localized region where protected attribute perturbations yield significantly divergent outcomes. 
While these $k$ inputs jointly expose a failure of fairness, existing explanation techniques fall short: differential debugging~\cite{10.1145/3368089.3409687,gaaloul2020mining} compares faulty and passing traces pairwise, and eXplainable AI (XAI) methods~\cite{ribeiro2016should,EthicalML-XAI, lundberg2017unified,yu2025fairlay} typically focus either on single-instance local explanations or on global approximations of the model.
To address this gap, we introduce a novel explanation framework tailored to discrimination clustering.
Our method leverages local perturbations and decision tree learning to uncover logical conditions that are shared across members of the $k$-discriminant set. Our experiments show its efficacy against the baseline XAI methods (RQ3). 
These explanations also help mitigate unfairness in the DNNs (RQ4).

\vspace{0.5em}\noindent \textbf{Contributions.} This paper makes the following contributions.
\begin{enumerate}[leftmargin=*]
    \item A novel fairness notion, $k$-discriminant, is introduced to characterize bias through structured clusters of discriminatory outcomes in deep neural networks.
    \item A \emph{mixed-integer linear programming} (MILP) based approach is developed to certify individual fairness and guide targeted randomized search.
    \item A novel \emph{explainable AI method} is proposed that leverages decision-tree learning and local perturbations to identify the root causes of $k$-discriminant clusters.
    \item \toolname, a hybrid framework, is presented that integrates formal verification, randomized search, and explanation techniques to detect, explain, and mitigate discrimination clustering in DNNs.
\end{enumerate}

\begin{figure*}
    \centering
    \includegraphics[width=1.0\textwidth]{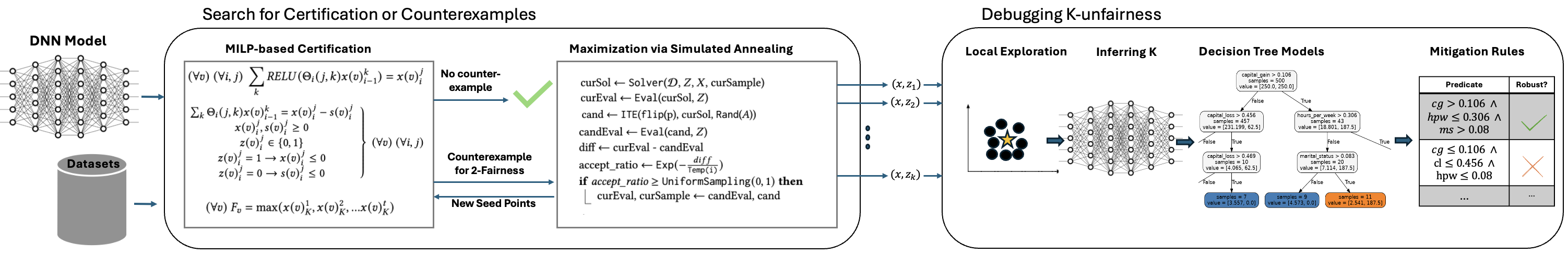}    
    \caption{\toolname Framework.}
    
    \label{fig:framework}
    \vspace{-1.0 em}
\end{figure*}

\section{Overview}
\label{sec:overview}

We focus on the notion of \textit{individual fairness}~\cite{dwork2012fairness}, where fairness is violated if altering protected attributes, while keeping all other features fixed, changes the machine learning outcome from favorable to unfavorable (or vice versa). 
Inspired by $\mathtt{DICE}$~\cite{Monjezi2023InformationTheoreticTA}, we define a model as $k$-discriminant (or $k$-unfair) if it produces $k$ distinct outcomes for a set of $K$ inputs ($k {\leq} K$) that differ only in their protected attributes.

\noindent \textbf{Real-world implications and advantages of k-discriminant.}
Our proposed generalization from the individual discrimination notion~\cite{angell2018themis} to k-discriminant acknowledges that fairness is rarely binary and provides a rigorous mathematical foundation for evaluating discrimination in complex socio-technological systems. 
One key implication is discovering significant ``arbitrariness'' in decision-making. This notion can reveal when k similar individuals receive significantly different likelihood scores, and hence it shows areas where the DNN's decision-making becomes highly arbitrary based on protected attributes.
Consider a loan application scenario where an unprivileged applicant is denied. When k=10, we found cases where ten applicants with nearly identical qualifications (e.g., credit scores within 1\% range, same income bracket, similar debt ratios) received vastly different loan approval probabilities based on their backgrounds. Individual fairness would only compare pairs and miss this systematic arbitrariness. By identifying areas of maximum arbitrariness, our explanation technique can pinpoint the specific combinations of features that trigger irrational decision-making.

\noindent \textbf{\toolname Workflow Summary.}
Figure~\ref{fig:framework} summarizes the workflow of \toolname. Given a dataset and a pre-trained DNN model, \toolname operates in two phases. In the \textit{search} phase, it encodes the DNN as a mixed-integer linear program (MILP) and verifies it against individual fairness. If a counterexample is found (a $2$-discriminant), \toolname initiates a randomized search—using both random walks and simulated annealing—to characterize the maximum $k$-discrimination starting from the counterexample.
\toolname alternates between verification and search, using the MILP solver to generate new random seeds for the search procedure, helping it escape potential plateaus in the input space.
For example, on the \emph{Adult Census Income} dataset~\cite{Dua:2019} (AC2 benchmark with $K = 90$), \toolname identified 18.6 ($\pm$ 6.7) $2$-discriminants on average, while \textsc{Fairfy}~\cite{10.1109/ICSE48619.2023.00134} found only 0.6 ($\pm$ 0.5). The average maximum $k$ was 17.8 ($\pm$ 0.4) for \toolname compared to 8.7 ($\pm$ 5.4) for \textsc{Fairfy}. Notably, only 3.8 ($\pm$ 2.3) of the 2,245 counterexamples exhibited the maximum $k$-discrimination of 19, highlighting the rarity and significance of these bugs.

In the \textit{debugging} step, given a set of inputs that exhibit maximum $k$-discrimination, we first explore their local neighborhood via random sampling and query the DNN to assess how $k$-discriminant behavior generalizes across nearby inputs. We then train a decision tree to explain the conditions under which the DNN exhibits clustered discrimination. The resulting model yields a set of logical predicates that evaluate to \texttt{true} when the DNN significantly discriminates.

To understand the causal influence of these predicates, we evaluate whether flipping the conditions in the explanation models can reduce the observed $k$—effectively identifying input subspaces where mitigating the predicate leads to fairer behavior. We then use these rules to implement guardrails that constrain the DNN’s behavior, and we leverage the associated test cases to fine-tune the model for mitigation.
Compared to \textsc{LIME}~\cite{ribeiro2016should}, \toolname provides more robust and succinct explanations with broader input coverage. On the AC-2 benchmark, our guided mitigation reduced the number of discriminatory instances from 2,245.0 to 769.6 and lowered the success rate of fairness bug reproduction from 87.7\% to 36.3\%.

\section{The Discrimination Clustering Problem}
\label{sec:definition}
In this study, we analyze deep neural network classifiers that have undergone prior training
(i.e., pre-trained DNN). 

\begin{definition}[DNN: Interpretation]
  A deep neural network (DNN) defines a function 
 \( F: X \times Z \rightarrow [0,1]^t \), where \( X = X_1 \times X_2 \times \ldots \times X_n \) denotes the space of non-protected input attributes (e.g., occupation, income, education), and \( Z = Z_1 \times Z_2 \times \ldots \times Z_m \) denotes the space of protected input attributes (e.g., race, gender, age).
  The output of $F$ is a \( t \)-dimensional probability vector over class labels. For any input
   \( (x, z) \), the predicted class label is given by
   \[
     F_{\text{label}}(x, z) =  \arg\max_{i \in [t]} F(x, z)(i).
   \]
We assume the domain of the protected attribute space 
$Z$ is finite, with $K$ denoting the number of distinct protected groups.
\end{definition}

\begin{definition}[DNN: Structure]
  A DNN \( F \) is defined by its input dimension \( n + m \), output dimension \( t \), number of hidden layers \( N \), and weight matrices \( \Theta_1, \Theta_2, \ldots, \Theta_N \). We analyze a pre-trained DNN with fixed parameters and weights to assess its fairness.
  For each layer \( r \in \{1, \ldots, N\} \), the output \( \mathcal{D}_r \) is computed as an affine transformation of the previous layer’s output \( \mathcal{D}_{r-1} \) using weights \( \Theta_r \), followed by a non-linear activation. Specifically:
  \begin{enumerate}
      \item For hidden layers \( 1 {\leq} r {<} N \), the activation is a ReLU function, defined as
      $\mathcal{D}_r = \max\{\Theta_r \cdot \mathcal{D}_{r-1}, 0\}$.
      
      \item For the output layer \( r = N \), a SoftMax function maps the final linear outputs to a probability distribution over the \( t \) classes.
  \end{enumerate}
  The term \( \mathcal{D}_r^j \) denotes the output of neuron \( j \) in layer \( r \).
  
\end{definition}

\noindent \textbf{The $2$-Discriminant Problem.}  
The prevailing notion of fairness, e.g., individual discrimination~\cite{10.1145/3510003.3510137,zhang2020white,9793943}, requires that similar individuals---who differ only in protected attributes---receive similar outcomes. Formally, a DNN is said to be \emph{$2$-discriminant} if there exist two protected attributes \( z_1 {\neq} z_2 \in Z \) and a shared unprotected input \( x {\in} X \) such that:
\[
\mathtt{Dist}_\epsilon\big(F(x, z_1), F(x, z_2)\big) > \epsilon,
\]
where \( \mathtt{Dist}_\epsilon \) measures the deviation between the outputs of the DNN for inputs differing only in protected attributes, and \( \epsilon \) is a specified tolerance threshold.
Conversely, we say the DNN satisfies \emph{individual fairness} if no such counterexample exists; that is:
$\forall x{\in}X, \forall z_1, z_2 {\in} Z, \mathtt{Dist}_\epsilon\big(F(x, z_1), F(x, z_2)\big){\leq} \epsilon.$
In this case, the DNN is said to be \emph{$2$-fair}, as no pair of counterfactuals differing only in protected attributes yields an output difference greater than \( \epsilon \).

\vspace{0.25em}
\noindent \textbf{Search for $k$-Discrimination.}  
From an AI risk perspective, we are often required to reason beyond $2$-discriminant behavior and assess the \emph{maximum unfairness} exhibited by a DNN model. We say a model is \emph{$k$-discriminant} if there exists a set of $K$ inputs  
$(x, z_1), (x, z_2), \ldots, (x, z_K)$,
where all records share the same non-protected features \( x \in X \) but differ in protected attributes \( z_i \in Z \), and the model produces \( k \leq K \) \emph{distinct} outputs. Formally:
\[
\exists z_1, \ldots, z_K {\in} Z,\ x {\in} X ~\text{s.t.}~ \texttt{C}_\epsilon\big(F(x, z_1),{\ldots}, F(x, z_K)\big){==} k,
\]
where \( \texttt{C}_\epsilon(\cdot) \) is a clustering function that partitions the outputs into \( 1 {\leq} k {\leq} K \) groups based on an indistinguishability threshold \( \epsilon \). Since the output of \( F \) lies in the range \([0, 1]\), the threshold \( \epsilon \) can be used to define uniform partitions of the outcome space for clustering.
We note that if the model is $2$-fair (i.e., not $2$-discriminant), then it is trivially $k$-fair for all \( k \geq 2 \). Otherwise, once a $2$-discriminant counterexample is found, we aim to compute the \emph{maximum value of $k$-discrimination} by solving:
\[
\max_{x \in X,\ z_1, \ldots, z_K \in Z}~\texttt{C}_\epsilon\big(F(x, z_1), \ldots, F(x, z_K)\big).
\]

To solve the $k$-discriminant problem, we construct up to \( k{+}1 \) copies of the DNN, denoted \( \mathcal{F} = F_1 \times \ldots \times F_{k+1} \), where each copy is evaluated on the same non-protected attributes but different protected attributes. Verifying fairness at level \( k \) involves checking whether the outputs of these \( k{+}1 \) copies can be clustered into fewer than \( k{+}1 \) distinct groups. Thus, increasing values of \( k \) require more copies of the model, making the verification problem increasingly expensive.

While formally verifying the absence of $k$-discrimination becomes computationally intractable as \( k \) grows, scalable techniques exist for certifying the $2$-discriminant property~\cite{10.1109/ICSE48619.2023.00134,Fischetti2018}. Our key insight is to leverage these solvers not just for certification, but also as a foundation to guide the quantification of the model's maximum $k$-discrimination.

\vspace{0.25em}
\noindent \textbf{Debugging for $k$-Discrimination.}  
We now turn to understanding the common conditions among edge-case inputs \( \mathcal{D} = \{(x, z_1), \ldots, (x, z_K)\} \) that exhibit significant discriminatory behavior in a DNN. Since this set collectively witnesses \( k \)-discrimination, standard methods such as differential debugging~\cite{10.1145/3368089.3409687,gaaloul2020mining} (comparing faulty vs. passing traces), local explanations~\cite{ribeiro2016should} (explaining a single instance), and global explanation techniques~\cite{EthicalML-XAI} (summarizing model behavior over the entire input space) fall short in explaining the root causes of such clustered discrimination.

Our goal is to identify conditions on non-sensitive attributes under which the model becomes disproportionately sensitive to protected attributes in its decision-making. The explanation challenge is to uncover what properties distinguish significantly discriminatory instances from benign ones.

\begin{tcolorbox}[boxrule=1pt,left=1pt,right=1pt,top=1pt,bottom=1pt]
\noindent \textbf{Discrimination Clustering Problem.} 
Given a pre-trained DNN model \( F \) with protected attributes \( Z \) (\( |Z| = K \)) and non-protected attributes \( X \), and a formal computational model of \( F \) that can certify \( 2 \)-fairness using a distance relation \( \mathtt{Dist}_\epsilon \), our goal is to:
\begin{enumerate}[label=(\roman*), leftmargin=2em]
    \item certify the DNN model \( F \) against \( 2 \)-fairness property,
    \item determine the maximum \( k \in [2, K] \) for which the model exhibits significant \( k \)-discrimination, and
    \item explain/mitigate the root causes of \( k \)-discrimination.
\end{enumerate}
\end{tcolorbox}

\section{\toolname{} for Discrimination Analysis}
\label{sec:approach}
We note that if no counterexample exists to the $2$-fairness property across all protected attributes, then the DNN is $k$-fair for all  $k {\geq} 2$. However, if a $2$-discriminant counterexample is found, the model violates individual fairness, and the goal is to characterize the extent of this violation by identifying the maximum value of $k$ or which the model is $k$-discriminant.

We address this in two phases. First, we formulate the $2$-fairness verification problem using symbolic reasoning techniques such as mixed-integer linear programming (MILP) and satisfiability modulo theories (SMT). These solvers can either certify the model's fairness or generate counterexamples that serve as individual discriminatory instances. Second, to assess the severity and structure of discrimination, we employ randomized search strategies that explore neighborhoods around these counterexamples to identify the maximum \( k \)-discrimination witnessed by the model.

\vspace{0.25 em}
\noindent \textbf{Certifying $2$-Fairness requirements.}
Unlike adversarial robustness, which is typically a local property around a specific input, individual fairness is a global property: any two inputs differing only in protected attributes---regardless of their location in the input space---must yield similar outputs. Therefore, verification approaches for local robustness~\cite{gopinath2018deepsafe,katz2017towards} are insufficient for certifying individual fairness.

\vspace{0.25 em}
\noindent \textit{SMT Solver.}
\textsc{Fairify}~\cite{10.1109/ICSE48619.2023.00134} formulates the individual
fairness problem with an SMT-based approach to verify individual fairness property in DNN. 
We briefly summarize this approach below. Let a DNN be viewed as a function $F: \mathbb{R}^m \rightarrow \mathbb{R}^n$ and let $F_1$ and $F_2$ be two copies of the same DNN, then any fairness property can be formulated as the verification query:
\[
\phi_{pre}(x, x') {\land}\phi_{dom}(x, x'){\land}y{=}F_1(x){\land} y'{=}F_2(x') {\rightarrow}\phi_{post}(y, y')
\]
where $x$ and $x'$ are the inputs to $F_1$ and $F_2$ respectively, $y$ and $y'$ are the outputs to $F_1$ and $F_2$ respectively, $\phi_{pre}$ is the precondition clause on the two inputs for the given fairness property (e.g., $x$ and $x'$ don't differ on the non-protected attributes), $\phi_{dom}$ is the domain constraints on the input and $\phi_{post}$ is the postcondition clause on the two outputs for the given fairness property (e.g., $y$ and $y'$ are equal). Such a verification query can be fed into an SMT solver to check for satisfiability, and furthermore, can also be asked to construct counterexamples if it is unsatisfiable.  

\vspace{0.25 em}
\noindent \textit{Mixed Integer Linear Programs.} We use a MILP encoding, similar
to~\cite{Fischetti2018,narodytska2018verifying,dutta2018output}
to certify individual fairness of DNN. 
We describe the MILP formulation of our DNN inspired by~\cite{Fischetti2018}. We have two copies of a pertained DNN each with $R$ hidden layers, input dimension $n + m$, and output dimension $t$. Let the first $n$ dimensions of the input be the non-protected input and the latter $m$ dimensions be the protected input. If $x(v)_r^j$ denotes the output of the $j$th neuron in the $r$th layer for the $v^{th}$ DNN and $\Theta_i$ denotes the weight for layer $i$ for both the DNNs, then the computation equations for our DNNs are:
\begin{align*}
    (\forall v) \: (\forall i,j) \: \sum_{r} \mathtt{ReLU}(\Theta_i(j,r) x(v)_{i-1}^r) = x(v)_i^j 
\end{align*}
We can denote these constraints in linear form by introducing new variables 
$s(v)_r^j$ and $z(v)_r^j$, which represent the negative output of the $j$th neuron in the $r$th layer and the activation state of the $\mathtt{ReLU}$ unit acting on the $j$th neuron in the $r$th layer respectively. Noting that $\mathtt{ReLU}(x) := \max (x , 0)$, we get our corresponding linear constraints to be: for all $ v, i$ and $j$: 
\begin{eqnarray*}
\sum_{r} \Theta_i(j,r) x(v)_{i-1}^r &=&x(v)^{j}_{i}-s(v)^{j}_{i} \\
x(v)^{j}_{i}, s(v)^{j}_{i} &\geq& 0 \\
z(v)^{j}_{i} &\in&\{0,1\} \\
z(v)^{j}_{i}=1 &\rightarrow& x(v)^{j}_{i} \leq 0 \\
z(v)^{j}_{i}=0 &\rightarrow& s(v)^{j}_{i} \leq 0 
\end{eqnarray*}
Additionally, our input to both DNNs is identical on the non-protected inputs, i.e.,
\begin{align*}
    (\forall j \le n) \: x(1)_0^j = x(2)_0^j
\end{align*}
for $j \in \{1,\ldots, n\}$.
If we denote the final output of the DNNs by $F_1$ and $F_2$ respectively, then $F_v$ is given by the maximum of the output of all neurons in layer $R$ i.e.
\begin{align*}
(\forall v) \: F_v = \max ( x(v)_R^1, x(v)_R^2, ... x(v)_R^t  )    
\end{align*}
As before we can denote these constraints in linear form by introducing auxiliary variables. Finally, our optimization objective is given by $\max | F_1 - F_2 |$.
Our objective function can be made linear by noting that $|x| = \max (x, -x)$ and then using the techniques shown before to convert the \textit{max} function into a linear form by introducing auxiliaries. 
Finally, it is trivial to see that our two DNNs satisfy the 2-Fairness property iff the output of the corresponding MILP is at most $\epsilon$ (we use $\epsilon$ of 0.05 similar to \textsc{Fairify}~\cite{10.1109/ICSE48619.2023.00134}).

\vspace{0.5em}
\noindent \textbf{Finding $k$-Discriminants with Randomized Search.}  
Algorithm~\ref{alg:overall-search} takes as input a deep learning model \( \mathcal{D} \); a dataset \( A \) consisting of protected attributes \( P \) and non-protected attributes \( Q \); a \texttt{Solver} (either \texttt{SMT} or \texttt{MILP}); an \texttt{Eval} function that measures the fitness of the solution against the $K$-fairness criterion; a search type \( \tau \) (e.g., random walk, simulated annealing (SA), or a hybrid SA with nearest-neighbor heuristics); a temperature function for SA; and a timeout \( T \). The algorithm outputs a solution that witnesses a violation of the $k$-fairness requirement.
Following \textsc{Fairify}~\cite{10.1109/ICSE48619.2023.00134}, we set \( \epsilon {=} 0.05 \), meaning the DNN's outputs are considered distinct if they differ by more than 5\% in predicted score as the protected attributes vary while others remain fixed.

\begin{algorithm}[t!]
	\DontPrintSemicolon
	\KwIn{Deep learning model $\Dd$, Test Data Samples $A$, 
	protected attributes $Z$, Non-Protected attributes $X$,
    Formal computation model \texttt{Solver},
    An evaluation function \texttt{Eval},
    Type of search $\tau$: `RW', `SA', or `SA+KNN',
	Temperature of SA temp,
    head probability p, and 
	Time-out $T$.
	}
	\KwOut{bestEval, bestidx}

    seed, $i$, bestEval $\gets$ $\mathtt{Rand}$($A$), 0, 0

    curSol $\gets$ $\mathtt{Solver}$($\Dd$, $Z$, $X$, seed)

    curEval $\gets$ $\mathtt{Eval}$(curSol, $Z$)
    
    \While{Not Time-out($T$)}{
        
        \If{curEval $>$ bestEval}{
            bestEval, bestSol $\gets$ curEval, curSol
        }
        \Else{
            curSol, curEval $\gets$ $\mathtt{Solver}$($\Dd$, $Z$, $X$, curSol), $\mathtt{Eval}$(curSol, $Z$)
        }

        \If{$\tau$ == RW}{
            cand $\gets$ $\mathtt{RandNeighbor}$($A$, curSol)

            curSol, curEval $\gets$ cand, $\mathtt{Eval}$(cand, $Z$)
        }
        \Else{
            \If{$\tau$ == SA}{
                cand $\gets$ $\mathtt{ITE}$($\mathtt{flip}$(p), curSol, $\mathtt{Rand}$($A$))
             }
             \ElseIf{$\tau$ == SA+KNN}{
                cand $\gets$ $\mathtt{ITE}$($\mathtt{flip}$(p), $\mathtt{KNN}$(curSol), $\mathtt{Rand}$($A$))        
             }

            candEval $\gets$ $\mathtt{Eval}$(cand, $Z$)
        
            diff $\gets$ curEval - candEval
                        
            accept\_ratio $\gets$ $\mathtt{Exp}(-\frac{diff}{\mathtt{Temp(i)}})$
            
            \If{accept\_ratio $\geq$ $\mathtt{UniformSampling}(0,1)$}{  
                curSol, curEval $\gets$ cand, candEval
            }
        }
        
        i $\gets$ i + 1
    }  
    \Return {(bestEval, bestIdx)}

\caption{\textsc{\toolname Search}}
\label{alg:overall-search}
\end{algorithm}

\begin{algorithm}[t!]
\DontPrintSemicolon
\KwIn{
    DNN model \( \mathcal{D} \);
    $k$-discriminant input set \( X = \{(x, z_1), \ldots, (x, z_K)\} \);
    number of neighborhood samples \( n \);
    high $k$-discrimination threshold \( \kappa \);
    significant difference threshold \( \delta \).
}
\KwOut{Explanation predicates \( \phi \)}

\BlankLine
\( X_{\text{perturb}} \gets \mathtt{localPerturbation}(X, n) \) 

{\scriptsize \tcp*{Generate neighborhood samples}}

\( \text{Scores} \gets \mathcal{D}(X_{\text{perturb}}) \) 
{\scriptsize \tcp*{Evaluate perturbed inputs}}

\( Y_{\text{perturb}} \gets \mathtt{getHighLowKLabels}(\text{Scores}, \kappa) \) 

{\scriptsize \tcp*{Label based on $k$}}

\( \Phi \gets \mathtt{buildDecisionTree}(X_{\text{perturb}}, Y_{\text{perturb}}) \) 

{\scriptsize\tcp*{Train decision tree}}

\( L, \Pi \gets \mathtt{getLeavesAndPaths}(\Phi) \) 

{\scriptsize\tcp*{Extract leaves and paths}}

\ForEach{\((l, \pi) \in (L, \Pi)\)}{
    \If{$\mathtt{isHighKLeaf}$($l$)}{
        \( X_{\text{cex}} \gets \mathtt{sampleCounterExamples}(\pi) \) \;

        \( \text{Scores}_{\text{cex}} \gets \mathcal{D}(X_{\text{cex}}) \) \;

        \( \text{meanKDiff} \gets \mathtt{getKStats}(\text{Scores}, \text{Scores}_{\text{cex}}) \) \;

        \If{\( \text{meanKDiff} \geq \delta \)}{
            \( \phi.\mathtt{add}(\pi) \) \;
        }
    }
}
\( \mathcal{D}' \gets \mathtt{train}(\mathcal{D}, X, \phi) \) 

{\scriptsize \tcp*{Retrain with explanation-based mitigation}}

\Return \( \phi, \mathcal{D}' \)

\caption{\textsc{\toolname Debugging}}
\label{alg:AlgorithmExplViaDT}
\end{algorithm}

\noindent \textit{Initialization}: A random data sample from the dataset is selected as the current seed to query \texttt{MILP} or \texttt{Z3} solvers. We then use the current seed to search the DNN $\Dd$ via the \texttt{Solver} that finds a $2$-discriminant. We take the instance and query the DNN over all possible protected values to measure the number of buckets for the evaluation of instances. 

\noindent \textit{Iterations}: The loop executes until the timeout \(T\) is reached. In each iteration, we first check if the current solution gives the best solution.
Otherwise, we might be stuck in the local minima, hence we query the \texttt{Solver} to get a new solution. 
Depending on the type of search (discussed below), we generate a candidate sample from the current solution (derived by the Solver) and accept it as the current sample. 

\noindent \textit{Search Type}: The logic of search significantly depends on the type of search $\tau$.
We consider the following search strategy:

\begin{enumerate}
    \item \textbf{Stateless Random Search}. We randomly choose a data point from the neighborhood of the current solution and accept it as the next sample to explore. Since this approach navigates the search space uniformly at random without any guidance; this serves as the baseline. 

    \item \textbf{Simulated Annealing Search}. We consider the current solution, inferred by the solver, 
    with the probability $p$ and a random data point from $A$ with the probability $1-p$ as the candidate
    for some large $p\geq 0.9$. Then,
    we follow the Metropolis algorithm, where we evaluate the fitness of the candidate and accept it as
    the next sample with the probability that is proportional to its difference from the fitness of the current solution.
    
    \item \textbf{Simulated Annealing with Nearest Neighbor Search}. This type of search uses nearest neighbors of current solutions in addition
    to a random sample from $A$. Specifically, the algorithm uniformly chooses one of the nearest neighbors with
    the probability $p$ and a random data point from $A$ with the probability $1{-}p$ as the candidate
    for some large $p{\geq} 0.9$.
\end{enumerate}

\vspace{0.5 em}
\noindent \textbf{Debugging $k$-discrimination.}
Given the set of $k$ samples that witness the maximum $k$-discrimination in the DNN, i.e., $\Dd=\set{(x,z_i)}_{i=1}^{K}$, our next goal is to understand the circumstances under which the DNN model becomes arbitrary and come up with a mitigation strategy. 
One naive idea is to infer $K$ local explanations via methods like \textsc{Lime}~\cite{ribeiro2016should} and mine common patterns among the $K$ explanatory models. However, this approach might be both expensive and fail to find useful patterns, and derive limited explanations. We need a robust framework to learn a common explanation for all $K$ points together. Our approach is to leverage the decision tree algorithms to synthesize a set of predicate functions
$\phi_j: X \to 	\mathbb{B}$ such that $\phi_j(x) = \texttt{true}$ provides an explanation about the circumstances over the non-protected attributes that led to the maximum arbitrariness of DNNs.

Algorithm~\ref{alg:AlgorithmExplViaDT} shows our approach to inferring the predicate
functions to explain bugs. Given a set of inputs that witnesses significant discrimination, we first generate $n$ neighborhood data samples around the input by local perturbations (Line 1). Then, we query the DNN model to measure $k$-discrimination for each of those neighborhood data samples (Line 2).
To provide a succinct explanation, we convert $k$ values into high and low labels (binary classes of high-K vs. low-K) using the 0.95-percentile of $k$ values, where only the top 5\% of $k$ values belong to the high-K class of significantly discriminatory (Line 3). 
Then, we infer a decision tree model that yields a set of paths (a predicate function) in the hyper-rectangular input space to explain what properties are common inside the high-class inputs and what properties distinguish high and low classes (Line 4). 
Then, we retrieve each path in the tree and its corresponding leaf node and go through each path and label to identify explanatory models (Line 5-6). If the leave node is a high-K class (Line 7), then the corresponding path in the tree that traverses from the root node to this leaf node is a candidate to explain the significant discriminatory inputs. To add the candidate path to the final set of explanatory models, we sample data points that evaluate the candidate path to \texttt{false}, i.e., negating the predicate function (Line 8), and compute the corresponding $k$ values of these instances by querying the DNN model (Line 9). We calculate the difference between the mean of $k$ for the data samples that satisfy the path conditions and those that negate the conditions (Line 10). If the differences are more than a threshold $\delta$, we deem the path robustly explains the significant discriminatory instances (Line 11-12). We use the generated samples and DT predicates to mitigate unfairness~\cite{dasu2024neufair} in the DNNs (Line 13).

\section{Experiments}
\label{sec:experiments}

In this paper, we pose the following research questions:

\begin{enumerate}[start=1,label={\bfseries RQ\arabic*},leftmargin=3em]

    \item How
    does \toolname compared to \textsc{Fairfy}~\cite{10.1109/ICSE48619.2023.00134} in certifying $2$-fairness and finding counterexamples? 
    
    \item What is the performance of different randomized search algorithms in characterizing $k$-discrimination?

    \item What are the performance and complexity of \toolname's explanations as compared to the baseline \textsc{LIME}~\cite{ribeiro2016should}?    

    \item How does \toolname help improve fairness?

\end{enumerate}

\vspace{0.25 em}
\noindent \textbf{Benchmarks.} 
We have 20 DNN benchmarks of various architectures (fully connected and based on ReLU activation functions) from the literature~\cite{10.1109/ICSE48619.2023.00134}. They include real-world DNNs in Kaggle \cite{10.1145/3368089.3409704}
and the SE literature~\cite{10.1145/3377811.3380331,udeshi2018automated,10.1145/3510003.3510137,10.1145/3428253, 10.1007/978-3-030-88806-0_15} where the number of layers and neurons vary from 3-11 and 10-318 neurons. These benchmarks are trained over two popular and socially critical datasets.
The Adult Census dataset concerns whether an individual earns more than 50K or not. The Bank Marketing dataset is used for the prediction of whether a client will subscribe to a service or not. 

\vspace{0.25 em}
\noindent \textbf{Technical Details.} We implemented \toolname in Python v3.8.10 with TensorFlow v2.12.0 (Keras API) and scikit-learn v1.2.2. 
We run all the experiments on an Ubuntu 20.04.4 LTS OS  sever with AMD Ryzen Threadripper PRO 3955WX 3.9GHz 16-cores X 32 CPU and two NVIDIA GeForce RTX 3090 GPUs.
We take the average of $10$ multiple runs for all experiments and report the standard deviations from the mean.
We also set $K$ to 20 throughout the experiments.  

\noindent \textit{Hyperparameter Selection.} Following \textsc{Fairify}~\cite{10.1109/ICSE48619.2023.00134}, we set the MILP solver timeout to 100 seconds and used the default convergence tolerance of 0.0001. We run the randomized search for 4 hours.
While we did not fix a threshold for k-discrimination during the search, we used a 95-percentile threshold to classify samples into high and low k values for training decision trees.  
We generated 5,000 samples following \textsc{LIME}~\cite{ribeiro2016should} established methodology, from the neighborhood of high k-value instances to explain circumstances under which the DNNs exhibited arbitrary behaviors. 
This number balanced the explanation quality with computational efficiency, required to quantify the impact of local samples on model outcomes. 
\begin{table*}[!tb]
\caption{Comparison of SMT Solver to MILP solver for characterizing $k$-fairness.
}
\centering
\resizebox{0.95\textwidth}{!}{%
\begin{tabu}{|l|c|c|c|c|c|c|c|c|}

\hline
\multirow{2}{2.5em}{DNN} & \multicolumn{4}{| c |}{\textsc{Fairfy}~\cite{10.1109/ICSE48619.2023.00134} (SMT)} & \multicolumn{4}{| c |}{\toolname (MILP)} \\\cline{2-9}

& \texttt{$\#ID$} & \texttt{$T.1st$} (s) & \texttt{$\#K.1st$} & \texttt{Num. Clusters} (\#K) & \texttt{$\#ID$} & \texttt{$T.1st$} (s) & \texttt{$\#K.1st$} & \texttt{Num. Clusters} (\#K) \\
\hline

AC1 & 1.4 ($\pm$ 1.1) &	81.5 ($\pm$ 70.2) &	8.2 ($\pm$ 4.9) &	11.2 ($\pm$ 5.1) &					\textbf{23.2} ($\pm$ 2.9) &	\textbf{0.5} ($\pm$ 0.2) &	\textbf{17.6} ($\pm$ 0.9) &	\textbf{18.6} ($\pm$ 0.6)  \\ 
\hline

AC2 & 0.6 ($\pm$ 0.5) &	75.2 ($\pm$ 108.5) &	8.7 ($\pm$ 5.4) &	8.7 ($\pm$ 5.4) & \textbf{18.6} ($\pm$ 6.7) &	\textbf{2.7} ($\pm$ 2.2) &	\textbf{15.8} ($\pm$ 1.1) &	\textbf{17.8} ($\pm$ 0.4)  \\ 
\hline

AC3 & 2.0 ($\pm$ 2.8) &	32.0 ($\pm$ 29.9) &	9.2 ($\pm$ 5.7) &	11.2 ($\pm$ 6.6) &	\textbf{12.6} ($\pm$ 4.6) &	\textbf{3.4} ($\pm$ 2.0) &	\textbf{11.4} ($\pm$ 3.0) &	\textbf{16.0} ($\pm$ 2.4) \\ 
\hline
AC4 & \textbf{0.4} ($\pm$ 0.9) &	\textbf{71.2} ($\pm$ 0.0) &	\textbf{9.0} ($\pm$ 6.4) &	\textbf{12.0} ($\pm$ 8.5) &	\textbf{0.4} ($\pm$ 0.5) &	106.2 ($\pm$ 50.3) &	8.5 ($\pm$ 2.1) &	8.5 ($\pm$ 2.1)  \\ 
\hline
AC5 & 0.0 ($\pm$ 0.0) &	N/A($\pm$ N/A) &	N/A($\pm$ N/A) &	N/A($\pm$ N/A) &	\textbf{4.0} ($\pm$ 0.7) &	\textbf{25.1} ($\pm$ 15.8) &	\textbf{10.2} ($\pm$ 4.4) &	\textbf{15.0} ($\pm$ 0.0)  \\ 
\hline
AC6 & 0.4 ($\pm$ 0.5) &	139.1 ($\pm$ 83.5) &	9.5 ($\pm$ 0.7) &	9.5 ($\pm$ 0.7) &	
\textbf{25.0} ($\pm$ 1.6) & \textbf{0.3} ($\pm$ 0.0) &	\textbf{11.8} ($\pm$ 4.4) &	\textbf{16.6} ($\pm$ 3.3)  \\ 
\hline
AC7 & 0.0 ($\pm$ 0.0) &	N/A($\pm$ N/A) &	N/A($\pm$ N/A) &	N/A($\pm$ N/A) &	\textbf{2.0} ($\pm$ 0.0) & \textbf{73.8} ($\pm$ 31.2) &	\textbf{15.2} ($\pm$ 3.3) &	\textbf{17.2} ($\pm$ 2.6)  \\ 
\hline
AC8 & 6.8 ($\pm$ 2.3) &	12.7 ($\pm$ 7.6) &	10.6 ($\pm$ 2.5) &	13.6 ($\pm$ 2.4) &
\textbf{25.0} ($\pm$ 0.7) &	\textbf{0.2} ($\pm$ 0.0) &	\textbf{15.8} ($\pm$ 0.8) &	\textbf{17.2} ($\pm$ 0.4)  \\ 
\hline
AC9 & 12.8 ($\pm$ 2.3) &	7.4 ($\pm$ 5.1) &	5.6 ($\pm$ 2.0) &	11.6 ($\pm$ 0.5) 
&	\textbf{23.6} ($\pm$ 1.3) &	\textbf{0.2} ($\pm$ 0.0) &	\textbf{14.8} ($\pm$ 0.8) &	\textbf{16.0} ($\pm$ 0.0)  \\ 
\hline
AC10 & 0.8 ($\pm$ 0.8) & 101.9 ($\pm$ 53.6)      & 6.7 ($\pm$ 5.7)           &	9.7 ($\pm$ 6.7) 
&	\textbf{23.4} ($\pm$ 1.1) &	\textbf{0.43} ($\pm$ 0.0) &	\textbf{15.0} ($\pm$ 1.6) &	\textbf{17.6} ($\pm$ 0.5)  \\ 
\hline
AC11 & 0.0 ($\pm$ 0.0) &	N/A($\pm$ N/A) &	N/A($\pm$ N/A) &	N/A($\pm$ N/A) 
&	\textbf{0.8 ($\pm$ 1.3)} & \textbf{64.2} ($\pm$ 53.9) &	\textbf{8.5} ($\pm$ 0.7) &	\textbf{9.0} ($\pm$ 1.4)  \\ 
\hline
AC12 & 0.0 ($\pm$ 0.0) &	N/A($\pm$ N/A) &	N/A($\pm$ N/A) &	N/A($\pm$ N/A) 
&	\textbf{13.2 ($\pm$ 4.4)} &	\textbf{3.8} ($\pm$ 2.9) &	\textbf{15.2} ($\pm$ 3.2) &	\textbf{18.8} ($\pm$ 1.6)  \\ 
\hline
BM1 & 1.2 ($\pm$ 1.6) &	4.19 ($\pm$ 0.1) &	3.0 ($\pm$ 0.0) &	6.0 ($\pm$ 4.24) 
&	\textbf{37.8} ($\pm$ 4.4) &	\textbf{2.4} ($\pm$ 1.4) &	\textbf{7.8} ($\pm$ 0.8) &	\textbf{9.0} ($\pm$ 0.0)  \\ 
\hline
BM2 & 5.0 ($\pm$ 6.4)          &	42.6 ($\pm$ 41.3) &	4.2 ($\pm$ 3.0)          &	4.4 ($\pm$ 3.3) 
& \textbf{33.2} ($\pm$ 11.5) &	\textbf{5.3} ($\pm$ 4.8)   &	\textbf{7.4} ($\pm$ 1.1) &	\textbf{9.0} ($\pm$ 0.0)   \\ 
\hline
BM3 & 4.4 ($\pm$ 4.0) &	22.3 ($\pm$ 46.7) &	4.0 ($\pm$ 2.3) &	5.2 ($\pm$ 2.5) 
    & \textbf{58.8} ($\pm$ 2.8) &	\textbf{2.3} ($\pm$ 1.1) &	\textbf{7.6} ($\pm$ 1.1) &	\textbf{9.0} ($\pm$ 0.0)  \\ 
\hline
BM4 & \textbf{0.6 ($\pm$ 1.3)} &	\textbf{6.5} ($\pm$ 0.0) &	\textbf{2.0} ($\pm$ 0.0) &	\textbf{2.0} ($\pm$ 0.0) 
&	0.0 ($\pm$ 0.0) &	N/A($\pm$ N/A) &	N/A($\pm$ N/A) &	N/A($\pm$ N/A)  \\ 
\hline
BM5 & 13.8 ($\pm$ 12.6) &	5.8 ($\pm$ 10.7) &	2.4 ($\pm$ 0.5) &	3.4 ($\pm$ 0.9) 
& \textbf{43.2} ($\pm$ 4.8) &	\textbf{1.6} ($\pm$ 1.2) &	\textbf{7.8} ($\pm$ 1.3) &	\textbf{9.0} ($\pm$ 0.0)  \\ 
\hline
BM6 & 17.2 ($\pm$ 14.1) &	17.7 ($\pm$ 37.5)       &	\textbf{5.2} ($\pm$ 2.5) &	\textbf{8.2} ($\pm$ 1.1) 
&	\textbf{59.8} ($\pm$ 6.7) &	\textbf{3.5} ($\pm$ 2.1) &	4.0 ($\pm$ 1.0) &	5.8 ($\pm$ 0.4)  \\ 
\hline
BM7 & \textbf{8.4} ($\pm$ 9.2) &	\textbf{24.1} ($\pm$ 50.3) &	\textbf{4.0} ($\pm$ 1.4) &	\textbf{5.2} ($\pm$ 1.3) 
&	0.0 ($\pm$ 0.0) &	N/A($\pm$ N/A) &	N/A($\pm$ N/A) &	N/A($\pm$ N/A) \\ 
\hline
BM8 & 0.0 ($\pm$ 0.0) &	N/A($\pm$ N/A) &	N/A($\pm$ N/A) &	N/A($\pm$ N/A) 
& \textbf{1.6} ($\pm$ 0.9) &	\textbf{63.7} ($\pm$ 40.4) &	\textbf{4.6} ($\pm$ 0.9) &	\textbf{4.8} ($\pm$ 0.8)  \\ 
\hline
\end{tabu}
}
\label{table:Z3-MILP}
\end{table*}

\noindent \textbf{RQ1: Finding individual discrimination.}
\label{sec:rq1}
We compare our tool to the \textsc{Fairify}~\cite{10.1109/ICSE48619.2023.00134}, a formal verification method to certify $2$-fairness properties of DNNs or find counterexamples as the individual discriminatory instances ($2$-discriminants). 
The key difference between \toolname and \textsc{Fairify} is that Fairify models the DNNs with SMT formulations and uses Z3-solver whereas \toolname models DNNs with MILP formulations. 

Table \ref{table:Z3-MILP} compares Fairify to our tool \toolname. The statistically significant results are highlighted with bold fonts in the tables.
Here, we report $\#ID$ - the number of individual discriminatory (ID) instances, $T. 1st~(s)$ - time to the first ID, $\#K.1st$ - $k$-discrimination for the first ID instance, and $\#K$ -  the maximum $k$-discrimination. 

We consider \#ID and \#K as the metrics for efficacy, and the T.1st as the metric for efficiency.
\toolname outperforms Fairify in 85\% of the cases (17 out of 20 cases) in finding more IDs in the given timeout. In 5 cases, Fairify does not find any ID instances where \toolname finds a good number of IDs.
Considering \#K, \toolname outperforms Fairify in 80\% of the cases.
In terms of efficiency of finding the first solution (i.e., T.1st), \toolname outperforms Fairify in 90\% of cases. Farify spends a significant amount of time on preprocessing (input partition) and pruning.  

We note that in the BM4 and BM8 cases, only one of the tools can find discriminatory instances. 
This leads us to understand crucial differences between MILP- vs. Z3-based techniques. 
We find that Z3 produces counterexamples that are often close to each other and almost belong to the same region. On the other hand, MILP produces counterexamples from different regions of the search space, oftentimes from the boundaries that separate a safe region from an unsafe one. Each technique might be effective for different benchmarks.

\begin{tcolorbox}[boxrule=1pt,left=1pt,right=1pt,top=1pt,bottom=1pt]
\textbf{Answer RQ1:} Our MILP formulation significantly outperforms the baseline Fairify in most of the cases (more than 80\% of the time) in terms of finding more individual discriminatory (ID) instances in shorter amounts of time. 
\end{tcolorbox}

\begin{table*}[!tbh]

\caption{Random Walk (RW) vs. Simulated Annealing (SA) as well as K-nearest neighbors variant of SA.}

\centering
\resizebox{0.97\textwidth}{!}{%

\begin{tabu}{|l|l|l|l|l|l|l|l|l|l|}
    \hline
        Model & Search & Iter($\times$ 1K) & $Max.K$ & $Avg.K$ & $T.1st$ & $\#ID$ & $Succ.rate$ & $\#ID.maxK.1$ &  $T.maxK.1$ \\ \hline
        \multirow{3}{2.5 em}{AC1} & RW & 2.7 ($\pm$ 0.4) & \textbf{20.0} ($\pm$ 0.0) & 8.1 ($\pm$ 0.0) & 5 ($\pm$ 10) & 1809 ($\pm$ 239) & \textbf{67.2} ($\pm$ 0.7) & \textbf{3.8} ($\pm$ 2.1) & 4561 ($\pm$ 6387) \\
        ~ & SA & \textbf{3.6} ($\pm$ 1.1) & \textbf{20.0} ($\pm$ 0.0) & \textbf{9.3} ($\pm$ 1.1) & \textbf{1} ($\pm$ 2) & \textbf{1927} ($\pm$ 553) & 53.3 ($\pm$ 1.8) & 2.7 ($\pm$ 1.4)  & \textbf{1162} ($\pm$ 1723) \\ 
        ~ & SA+KNN & 2.1 ($\pm$ 0.3) & 19.3 ($\pm$ 0.5) & 8.8 ($\pm$ 0.2) & 62 ($\pm$ 86) & 625 ($\pm$ 200) & 29.7 ($\pm$ 9.9) & 1.9 ($\pm$ 0.7) &  4799 ($\pm$ 4089) \\ 
        \hline
        
        \multirow{3}{2.5 em}{AC2} & RW & 2.0 ($\pm$ 0.4) & 18.8 ($\pm$ 0.4) & 14.2 ($\pm$ 0.0) & 3 ($\pm$ 3) & 1953 ($\pm$ 366) & \textbf{97.4} ($\pm$ 0.2) & 2.0 ($\pm$ 2.2) & 6761 ($\pm$ 7330) \\ 
        
        ~ & SA & \textbf{2.6}($\pm$ 0.4) & \textbf{19.0} ($\pm$ 0.0) & 14.0($\pm$ 0.8) & \textbf{2} ($\pm$ 1) & \textbf{2245} ($\pm$ 312) & 87.7 ($\pm$ 0.6) & 3.8 ($\pm$ 2.3) &  4769 ($\pm$ 3939) \\ 
        ~ & SA+KNN & 1.3($\pm$ 0.3) & 18.3 ($\pm$ 0.5) & \textbf{14.3} ($\pm$ 0.1) & 8 ($\pm$ 2) & 885($\pm$ 149) & 71.7 ($\pm$ 14.0) & \textbf{4.0} ($\pm$ 2.5) & \textbf{3328} ($\pm$ 3733) \\ 
        \hline
        
        \multirow{3}{2.5 em}{AC3} & RW & 2.6($\pm$ 0.5) & \textbf{20.0} ($\pm$ 0.0) & 14.1 ($\pm$ 0.1) & \textbf{4} ($\pm$ 6) & 1667 ($\pm$ 310) & \textbf{64.3} ($\pm$ 0.5) & \textbf{255.2} ($\pm$ 51.7) &  \textbf{63} ($\pm$ 58) \\ 
        
        ~ & SA & \textbf{3.2} ($\pm$ 0.5) & \textbf{20.0} ($\pm$ 0.0) & 14.1 ($\pm$ 2.2) & \textbf{4} ($\pm$ 7) & \textbf{1846} ($\pm$ 353) & 58.6 ($\pm$ 6.3) & 170.4 ($\pm$ 72.0) &  86 ($\pm$ 80) \\
        
        ~ & SA+KNN & 2.0 ($\pm$ 0.5) & \textbf{20.0} ($\pm$ 0.0) & \textbf{17.5} ($\pm$ 0.5) & 20 ($\pm$ 20) & 105($\pm$ 32) & 5.8 ($\pm$ 2.9) & 12.5 ($\pm$ 6.2) & 808 ($\pm$ 1504) \\
        \hline
        
        \multirow{3}{2.5 em}{AC4} & RW & \textbf{0.2} ($\pm$ 0.0) & \textbf{20.0} ($\pm$ 0.0) & 15.2 ($\pm$ 0.4) & \textbf{122} ($\pm$ 66) & 122 ($\pm$ 4) & 88.0 ($\pm$ 3.3) & 4.9 ($\pm$ 1.7) &  2369 ($\pm$ 2335) \\ 
        
        ~ & SA & \textbf{0.2} ($\pm$ 0.0) & \textbf{20.0} ($\pm$ 0.0) & \textbf{17.1} ($\pm$ 0.4) & 127 ($\pm$ 47) & \textbf{125}($\pm$ 7) & \textbf{91.0} ($\pm$ 4.7) & \textbf{13.2} ($\pm$ 4.9) &  \textbf{1262} ($\pm$ 642) \\ 
        
        ~ & SA+KNN & 0.1($\pm$ 0.0) & 19.7 ($\pm$ 0.5) & 16.2 ($\pm$ 1.1) & 229 ($\pm$ 151) & 54 ($\pm$ 15) & 72.6 ($\pm$ 22.7) & 4.7 ($\pm$ 6.7) & 2841 ($\pm$ 2781) \\ 
        
        \hline
        
        \multirow{3}{2.5 em}{AC5} & RW & \textbf{1.7} ($\pm$ 0.3) & 18.1 ($\pm$ 0.5) & \textbf{13.7} ($\pm$ 0.0) & 4 ($\pm$ 3) & \textbf{1650} ($\pm$ 333) & \textbf{97.9} ($\pm$ 0.3) & 1.8 ($\pm$ 1.0) &  4730 ($\pm$ 6327) \\ 
        
        ~ & SA & 1.6($\pm$ 0.2) & \textbf{18.3} ($\pm$ 0.4) & 13.0 ($\pm$ 0.1) & \textbf{3} ($\pm$ 1) & 1545 ($\pm$ 235) & 97.7 ($\pm$ 0.76) & \textbf{9.6} ($\pm$ 5.7) &  \textbf{2209} ($\pm$ 2826) \\ 
        
        ~ & SA+KNN & 0.9 ($\pm$ 0.2) & 17.3 ($\pm$ 0.7) & 13.5 ($\pm$ 0.2) & 11 ($\pm$ 4) & 687 ($\pm$ 176) & 78.4 ($\pm$ 12.1) & 6.0 ($\pm$ 13.8) &  6483 ($\pm$ 4482) \\ 
        
        \hline
        
        \multirow{3}{2.5 em}{AC6} & RW & 2.6 ($\pm$ 0.5) & \textbf{20.0} ($\pm$ 0.0) & 14.3 ($\pm$ 0.0) & \textbf{1} ($\pm$ 0) & 2491 ($\pm$ 473) & \textbf{94.2} ($\pm$ 0.3) & 37.9 ($\pm$ 6.2) &  338 ($\pm$ 230) \\ 
        
        ~ & SA & \textbf{2.9} ($\pm$ 0.6) & \textbf{20.0} ($\pm$ 0.0) & 13.8 ($\pm$ 0.3) & \textbf{1} ($\pm$ 1) & \textbf{2623} ($\pm$ 500) & 91.5 ($\pm$ 1.5) & \textbf{108.4} ($\pm$ 25.7) &  \textbf{93} ($\pm$ 82) \\ 
        
        ~ & SA+KNN & 1.8 ($\pm$ 0.4) & \textbf{20.0} ($\pm$ 0.0) & \textbf{14.6} ($\pm$ 0.3) & 6 ($\pm$ 1) & 801 ($\pm$ 197) & 45.4 ($\pm$ 10.8) & 12.8 ($\pm$ 3.9) & 1847 ($\pm$ 2049) \\ 
        
        \hline
        
        \multirow{3}{2.5 em}{AC7} & RW & 4.1 ($\pm$ 0.8) & 16.0 ($\pm$ 0.0) & 14.4 ($\pm$ 0.3) & \textbf{820} ($\pm$ 770) & 17 ($\pm$ 3) & \textbf{0.4} ($\pm$ 0.1) & \textbf{7.1} ($\pm$ 1.8) & \textbf{3493} ($\pm$ 2993) \\
        
        ~ & SA & \textbf{4.3} ($\pm$ 0.9) & \textbf{16.5} ($\pm$ 0.5) & 14.3 ($\pm$ 0.4) & 1602($\pm$ 1527) & \textbf{18} ($\pm$ 7) & \textbf{0.4} ($\pm$ 0.2) & 2.6 ($\pm$ 1.6) &  3646 ($\pm$ 2413) \\ 
        
        ~ & SA+KNN & 3.8 ($\pm$ 1.0) & 16.0 ($\pm$ 0.5) & \textbf{15.0} ($\pm$ 1.0) & 2019 ($\pm$ 1579) & 7 ($\pm$ 3) & 0.2 ($\pm$ 0.1) & 2.9 ($\pm$ 1.8) &  7189 ($\pm$ 4688) \\
        
        \hline
        
        \multirow{3}{2.5 em}{AC8} & RW & 2.5 ($\pm$ 0.6) & 16.0 ($\pm$ 0.0) & 11.2 ($\pm$ 0.0) & 11($\pm$ 13) & 1467 ($\pm$ 370) & \textbf{58.5} ($\pm$ 0.8) & 7.5 ($\pm$ 3.3) &  \textbf{1804} ($\pm$ 1785) \\ 
        
        ~ & SA & \textbf{3.8} ($\pm$ 1.5) & \textbf{16.9} ($\pm$ 0.3) & \textbf{12.3} ($\pm$ 0.2) & \textbf{1} ($\pm$ 2) & \textbf{2072} ($\pm$ 518) & 57.0 ($\pm$ 8.8) & \textbf{9.2} ($\pm$ 3.6) &  2569 ($\pm$ 2182) \\
        
        ~ & SA+KNN & 2.2($\pm$ 0.4) & 15.6 ($\pm$ 0.7) & 12.0 ($\pm$ 0.2) & 12 ($\pm$ 9) & 222 ($\pm$ 48) & 10.3 ($\pm$ 2.1) & 5.0 ($\pm$ 4.6) & 3061 ($\pm$ 3627) \\ 
        \hline
        
        \multirow{3}{2.5 em}{AC9} & RW & 2.3 ($\pm$ 0.4) & \textbf{19.0} ($\pm$ 0.0) & \textbf{14.1} ($\pm$ 0.0) & 2 ($\pm$ 3) & 2080 ($\pm$ 343) & \textbf{91.3} ($\pm$ 2.9) & \textbf{53.3} ($\pm$ 8.3) &  \textbf{102} ($\pm$ 41) \\ 
        
        ~ & SA & \textbf{3.1} ($\pm$ 0.4) & \textbf{19.0} ($\pm$ 0.0) & 13.7 ($\pm$ 0.7) & \textbf{1} ($\pm$ 0) & \textbf{2430} ($\pm$ 323) & 77.6 ($\pm$ 1.4) & 51.0 ($\pm$ 8.9) &  443 ($\pm$ 362) \\ 
        
        ~ & SA+KNN & 2.3 ($\pm$ 0.3) & \textbf{19.0} ($\pm$ 0.0) & 13.8 ($\pm$ 0.2) & 9 ($\pm$ 6) & 443 ($\pm$ 183) & 19.7 ($\pm$ 9.7) & 5.2 ($\pm$ 1.5) &  1263 ($\pm$ 1258) \\ 
        \hline
        
        \multirow{3}{2.5 em}{AC10} & RW & 2.3($\pm$ 0.4) & \textbf{17.0} ($\pm$ 0.0) & 11.9 ($\pm$ 0.0) & \textbf{3} ($\pm$ 8) & 1795 ($\pm$ 311) & \textbf{77.1} ($\pm$ 0.5) & 1.8 ($\pm$ 0.9) &  4245 ($\pm$ 5628) \\ 
        
        ~ & SA & \textbf{3.5} ($\pm$ 0.9) & 16.9 ($\pm$ 0.3) & 11.3 ($\pm$ 1.6) & \textbf{3} ($\pm$ 4) & \textbf{1838} ($\pm$ 545) & 51.6 ($\pm$ 2.1) & \textbf{5.8} ($\pm$ 11.5) & \textbf{3778} ($\pm$ 3017) \\
        
        ~ & SA+KNN & 2.3 ($\pm$ 0.3) & 16.4 ($\pm$ 0.5) & \textbf{13.0} ($\pm$ 0.1) & 15 ($\pm$ 18) & 336 ($\pm$ 62) & 14.8 ($\pm$ 3.7) & 4.5 ($\pm$ 3.9) &  4779 ($\pm$ 4621) \\ 
        
         \hline
         
        \multirow{3}{2.5 em}{AC11} & RW & 1.1 ($\pm$ 0.3) & \textbf{20.0} ($\pm$ 0.0) & 13.8 ($\pm$ 0.1) & 10 ($\pm$ 6) & 89 ($\pm$ 275) & \textbf{81.2} ($\pm$ 1.3) & 7.2 ($\pm$ 3.1) &  2990 ($\pm$ 2735) \\
        
        ~ & SA & \textbf{1.3} ($\pm$ 0.3) & \textbf{20.0} ($\pm$ 0.0) & \textbf{15.0} ($\pm$ 0.4) & \textbf{8} ($\pm$ 5) & \textbf{987} ($\pm$ 275) & 73.8 ($\pm$ 7.1) & \textbf{29.9} ($\pm$ 16.2) &  \textbf{471} ($\pm$ 465) \\ 
        
        ~ & SA+KNN & 0.6 ($\pm$ 0.2) & 19.4 ($\pm$ 0.5) & 14.3 ($\pm$ 0.6) & 44 ($\pm$ 57) & 155 ($\pm$ 80) & 26.1 ($\pm$ 12.1) & 5.1 ($\pm$ 4.1) &  2498 ($\pm$ 2158) \\ 
        \hline
        
        \multirow{3}{2.5 em}{AC12} & RW & 2.2($\pm$ 0.4) & \textbf{20.0} ($\pm$ 0.0) & \textbf{17.7} ($\pm$ 0.1) & \textbf{1} ($\pm$ 1) & \textbf{2204} ($\pm$ 409) & \textbf{99.2} ($\pm$ 0.2) & \textbf{99.7} ($\pm$ 24.8) & \textbf{120} ($\pm$ 26) \\ 
        
        ~ & SA & \textbf{2.4} ($\pm$ 0.9) & \textbf{20.0} ($\pm$ 0.0) & 15.7 ($\pm$ 1.7) & \textbf{1} ($\pm$ 0) & 1662 ($\pm$ 593) & 70.8 ($\pm$ 1.6) & 60.6 ($\pm$ 25.6) & 203 ($\pm$ 149) \\ 
        
        ~ & SA+KNN & 1.5 ($\pm$ 0.1) & \textbf{20.0} ($\pm$ 0.0) & \textbf{17.7} ($\pm$ 0.1) & 9 ($\pm$ 4) & 957 ($\pm$ 185) & 65.5 ($\pm$ 11.0) & 19.6 ($\pm$ 8.0) & 2068 ($\pm$ 1431) \\ 
        \hline

        \multirow{3}{2.5 em}{BM1} & RW    & 8.8 ($\pm$ 1.9) & \textbf{9.0} ($\pm$ 0.0) &  5.0 ($\pm$ 0.0)	& 3 ($\pm$ 4)	& \textbf{5409} ($\pm$ 1151)	& \textbf{61.2} ($\pm$ 0.3) & \textbf{45.8} ($\pm$ 10.2)	& \textbf{295} ($\pm$ 43) \\
        ~                         & SA     & \textbf{9.1} ($\pm$ 1.6)  & \textbf{9.0} ($\pm$ 0.0) &  5.2 ($\pm$ 0.6)	& \textbf{1} ($\pm$ 1)	& 5135 ($\pm$ 657)	& 57.3 ($\pm$ 6.1) & 26.3 ($\pm$ 8.5)	&  420 ($\pm$ 393) \\ 
        ~                         & SA+KNN & 4.7($\pm$ 0.9)	 & \textbf{9.0} ($\pm$ 0.0) &  \textbf{5.7} ($\pm$ 0.1)	& 9 ($\pm$ 11)	& 1663 ($\pm$ 387)	& 35.3 ($\pm$ 3.8) & 	13.4($\pm$ 5.4)	& 1646 ($\pm$ 1018) \\ 
        \hline

        \multirow{3}{2.5 em}{BM2} & RW    & 11.0  ($\pm$ 0.8) & \textbf{9.0} ($\pm$ 0.0) & 4.0 ($\pm$ 0.0) & 12 ($\pm$ 13) & \textbf{766} ($\pm$ 53)	& \textbf{7.0} ($\pm$ 0.1)	& \textbf{22.2} ($\pm$ 2.0)	&  	\textbf{738} ($\pm$ 682) \\
        ~                         & SA     & \textbf{11.3} ($\pm$ 0.0)  & \textbf{9.0} ($\pm$ 0.0) & 4.6 ($\pm$ 1.1) & \textbf{8} ($\pm$ 10)  & 726 ($\pm$ 200) & 6.4 ($\pm$ 1.8)	& 17.3 ($\pm$ 7.0)	&  936 ($\pm$ 896) \\
        ~                         & SA+KNN & 9.0 ($\pm$ 1.8)   & \textbf{9.0} ($\pm$ 0.0) & \textbf{5.5} ($\pm$ 0.1) & 32 ($\pm$ 28) & 417 ($\pm$ 94)	& 4.6 ($\pm$ 0.2)	& 7.9 ($\pm$ 3.3)	&  3248 ($\pm$ 3087) \\
        \hline

        \multirow{3}{2.5 em}{BM3} & RW    & 10.6 ($\pm$ 1.8) & \textbf{9.0} ($\pm$ 0.0)	& 6.0 ($\pm$ 0.0) & 3 ($\pm$ 3)	  & \textbf{3230} ($\pm$ 551) & \textbf{30.3} ($\pm$ 0.0)	& 341.4 ($\pm$ 57.1) &  79 ($\pm$ 56)\\
        ~                         & SA     & \textbf{11.3} ($\pm$ 0.1) & \textbf{9.0} ($\pm$ 0.0)	& 6.5 ($\pm$ 0.9) & \textbf{2} ($\pm$ 2)	  & 2984 ($\pm$ 315) & 26.5 ($\pm$ 2.9)	& \textbf{419.1} ($\pm$ 132.0) &  \textbf{24} ($\pm$ 17)\\
        ~                         & SA+KNN & 7.9 ($\pm$ 2.2) & \textbf{9.0} ($\pm$ 0.0)	& \textbf{7.3} ($\pm$ 0.1) & 21 ($\pm$ 36) & 1072 ($\pm$ 276) & 13.6 ($\pm$ 0.7)	& 111.1 ($\pm$ 40.3) &  131 ($\pm$ 148)\\
        \hline

        \multirow{3}{2.5 em}{BM5} & RW    & \textbf{11.0} ($\pm$ 0.9) & \textbf{6.0} ($\pm$ 0.0) & 3.6 ($\pm$ 0.0) & \textbf{2} ($\pm$ 2)  & \textbf{2101} ($\pm$ 182) & \textbf{19.2} ($\pm$ 0.1) & 5.9  ($\pm$ 0.3)  &  1620 ($\pm$ 1549)\\
        ~                         & SA     & 10.2 ($\pm$ 1.7) & \textbf{6.0} ($\pm$ 0.0) & 3.7 ($\pm$ 0.5) & 3 ($\pm$ 3)  & 1605 ($\pm$ 249) & 15.9 ($\pm$ 2.5) & 25.4 ($\pm$ 22.2) & \textbf{464} ($\pm$ 2056)\\
        ~                         & SA+KNN & 10.7 ($\pm$ 1.6) & 5.9 ($\pm$ 0.3) & \textbf{4.1} ($\pm$ 0.0) & 11 ($\pm$ 9) & 1366 ($\pm$ 213) & 12.8 ($\pm$ 0.4) & \textbf{29.2} ($\pm$ 81.1) &  4836 ($\pm$ 2269)\\
        \hline

        \multirow{3}{2.5 em}{BM6} & RW    & \textbf{11.3} ($\pm$ 0.0) & 7.0 ($\pm$ 0.0) & 2.2 ($\pm$ 0.0) & 1.7 ($\pm$ 1.7) & \textbf{2450} ($\pm$ 0) &\textbf{21.7} ($\pm$ 0.0) & 4.0 ($\pm$ 0.0) &  1002 ($\pm$ 732)\\
        ~                         & SA     & \textbf{11.3} ($\pm$ 0.0) & \textbf{8.0} ($\pm$ 0.0) & \textbf{3.4} ($\pm$ 0.0) & \textbf{1.6} ($\pm$ 1.5) & 2315 ($\pm$ 328) & 20.5 ($\pm$ 2.9) & \textbf{21.5} ($\pm$ 13.1) &  \textbf{137} ($\pm$ 32)\\
        ~                         & SA+KNN & \textbf{11.3} ($\pm$ 0.1) & 7.9 ($\pm$ 0.3) & 2.8 ($\pm$ 0.1) & 5.0 ($\pm$ 4.1) & 2123 ($\pm$ 238) & 18.8 ($\pm$ 2.1) & 3.4 ($\pm$ 3.6) & 3239 ($\pm$ 1900)\\
        \hline

        \multirow{3}{2.5 em}{BM8} & RW    & \textbf{0.3} ($\pm$ 0.1) & 7.0 ($\pm$ 0.0) & 3.8 ($\pm$ 0.1) & \textbf{129} ($\pm$ 71)    & \textbf{52} ($\pm$ 17) & \textbf{18.5} ($\pm$ 2.6) & \textbf{1.2} ($\pm$ 0.5) & 4977 ($\pm$ 5861)\\
        
            ~   & SA     & \textbf{0.3} ($\pm$ 0.1) & \textbf{8.0} ($\pm$ 0.0)  & \textbf{4.8} ($\pm$ 0.1) & 394 ($\pm$ 87)   & 42 ($\pm$ 4)  & 16.5 ($\pm$ 3.5)  & \textbf{1.2} ($\pm$ 2.1) & \textbf{4570} ($\pm$ 5023)
        \\

            ~  & SA+KNN & 0.1 ($\pm$ 0.0) & 6.5 ($\pm$ 1.7) & 4.3 ($\pm$ 0.3) & 3415 ($\pm$ 478) & 20 ($\pm$ 9)   & 12.7 ($\pm$ 4.0)  & 2.0 ($\pm$ 2.0) & 8929.7 ($\pm$ 4458)
        \\
        \hline

\end{tabu}
}
\label{table:discriminatory}
\end{table*}

\noindent \textbf{RQ2: Search for $k$-discrimination.}
\label{sec:rq2}
Our goal is to find instances that witness the maximum $k$-discrimination. 
We compare the efficacy and efficiency of two variations of simulated annealing (SA) algorithms in characterizing the amounts of dependencies between protected attributes and DNN outcomes. We also include a Random Walk (RW) to establish a baseline that abolishes the SA.

Table~\ref{table:discriminatory} summarizes the outcomes of this experiment. Since we certify 2-fairness for BM4 and BM7, we exclude them from the search.
The columns are similar to Table~\ref{table:Z3-MILP}, except that $Iter$ shows the number of completed iterations, $succ.rate$ shows the proportion of ID instances relative to the total generated instances during the search, and $\#ID.maxK$ reports the number of IDs with max $k$-discrimination, and $T.maxK$ shows the time taken to find the max $k$-discrimination. 

To compare the performance of the search algorithms, we focus on the following metrics: $Max.K$, $Avg.K$, $\#ID.maxK$, and $\#T.maxK$. 
While all three randomized search algorithms effectively identify ID instances, the SA algorithm consistently outperforms the others, identifying ID instances with the maximum buckets in 94\% of the cases. 
When considering $\#ID.maxK$ and $T.maxK$, SA algorithms perform slightly better in 50\% of the cases, whereas RW and SA+KNN show better results in 39\% and 10\% of the cases, respectively. 
Also, taking into consideration the number of ID instances, the SA algorithm leads, finding the highest number of IDs in 56\% of the benchmarks, slightly surpassing the RW algorithm, which achieves this in 50\% of cases. 
Additionally, the SA algorithm demonstrates a clear advantage in terms of time to the first solution, discovering the first ID instance faster than the other algorithms in 72\% of the benchmarks. 
Focusing on the average number of clusters ($Avg.K$), the SA+KNN algorithm outperforms other algorithms by generating ID instances that exhibit a higher average number of clusters in 56\% of the benchmarks, where SA and SW algorithms achieve 33\% and 17\%, respectively.

\begin{tcolorbox}[boxrule=1pt,left=1pt,right=1pt,top=1pt,bottom=1pt]
\textbf{Answer RQ2:} Simulated annealing outperforms other randomized search strategies in inding maximum $k$-discrimination, the time to the first instance of max $k$, and the number of unique instances with max $k$-discrimination.
\end{tcolorbox}

\begin{table*}[!tb]
\caption{LIME~\cite{ribeiro2016should} vs. \toolname based on robustness, size, and the generality of explanation. 
}
\centering
\resizebox{0.75\textwidth}{!}{%
\begin{tabu}{|l|c|c|c|c|c|c|c|c|c|c|c|}
\hline
\multirow{2}{2.5em}{DNN} & \multirow{2}{2.5em}{\texttt{$Init.K$}} &\multicolumn{5}{| c |}{\textsc{LIME}~\cite{ribeiro2016should}} & \multicolumn{5}{| c |}{\toolname} \\\cline{3-12}

 & & \texttt{$Size$} & \texttt{$K$}& \texttt{$Pert.K$} & \texttt{$Diff$} & \texttt{Cov} & \texttt{$Size$} & \texttt{$K $} & \texttt{$Pert.K$} & \texttt{$Diff$} & \texttt{Cov} \\
\hline
AC1  & 20 & 5          & 20 & 8.5 ( ± 3.3)  & 11.5          & 1 & 6          & 12.7 (± 6.2 )                          & 1.1 (± 0.3)   & \textbf{11.6} & \textbf{108.2 (±8.0)}   \\ \hline
AC2  & 19 & 6          & 19 & 9.9 ( ± 3.4)  & 9.1           & 1 & 7          & 11.7 (± 3.0 )                          & 2.5 (± 2.5)   & \textbf{9.2}           & \textbf{15.4 (±3.7)}    \\ \hline
AC3  & 20 & 5          & 20 & 11 (± 4.3)    & 9.8           & 1 & 5          & 15.8 (±  2.0 )                         & 4.3 (± 4.5)  & \textbf{11.5} & \textbf{12.0 (±2.2)}    \\ \hline
AC4  & 20 & 7          & 20 & 16.1 (± 3)    & 3.9           & 1 & \textbf{3} & 19.5 (±  0.7 )                         & 8.2 (± 5.9)  & \textbf{11.3} & \textbf{384.7 (±23.7)}  \\ \hline
AC5  & 18 & 4          & 18 & 11 ( ± 0.7)   & 7.0             & 1 & \textbf{3} & 14.0 (± 0.0 )                           & 6.9 (± 3.2)  & \textbf{7.1}  & \textbf{1572.0 (±34.9)} \\ \hline
AC6  & 20 & 4          & 20 & 17.9 (± 0.8) & 2.3           & 1 & \textbf{3} & 17.0 (± 0.0 )                           & 8.2 (± 2.6)  & \textbf{8.8}  & \textbf{622.3 (± 22.8)}  \\ \hline
AC7  & 17 & 13         & 17 & 12 (± 3.6)    & 5.0             & 1 & \textbf{3} & 17.0 (± 0.0 )                           & 1.0 (± 0.0)  & \textbf{16.0}   & 1.0 (± 0.0)              \\ \hline
AC8  & 17 & 5          & 17 & 9.4 (± 1.4)   & 7.6           & 1 & 6          & \cellcolor[HTML]{F4CCCC}11.2 (±  0.9 ) & 1.2 (± 0.6)  & \textbf{10.0}   & \textbf{57.4 (± 8.9)}    \\ \hline
AC9  & 19 & \textbf{3} & 19 & 6.7 (± 1.7)   & \textbf{12.3} & 1 & 4          & 16.0 (±  0.0)                           & 4.0(±  0.0)  & 12.0  & \textbf{493.0 (± 25.8)}  \\ \hline
AC10 & 17 & 5          & 17 & 7.9 (± 4.3)   & \textbf{9.1}  & 1 & 3          & 17.0 (±  0.0 )                           & 10.0 (± 0.0) & 7.0             & \textbf{2590.5 (± 55.4)} \\ \hline
AC11 & 20 & 6          & 20 & 16 (± 1.7)    & 4.2           & 1 & \textbf{4} & 13.0 (± 1.4 )                         & 4.3 (± 2.1)  & \textbf{8.7}  & \textbf{895.3 (± 27.5)}  \\ \hline
AC12 & 20 & 6          & 20 & 10.8 (± 2.7)  & 9.2           & 1 & \textbf{4} & 19.0 (± 1.4 )                         & 1.8(± 1.2)  & \textbf{17.2} & \textbf{702.9 (± 21.7)}  \\ \hline
BM1  & 9  & 3          & 9  & 3.7 (± 1.0)     & 5.3           & 1 & 4          & 9.0 (± 0.0 )                          & 1.9 (± 2.2)  & \textbf{7.1}  & \textbf{77.5 (± 7.9)}    \\ \hline
BM2  & 9  & 4          & 9  & 5.3 (± 0.6)   & 3.7           & 1 & 8          & 5.2 (±  3.2 )                          & 1.2 (± 0.6)  & \textbf{4.0}    & \textbf{19.1 (± 4.6)}    \\ \hline
BM3  & 9  & 3          & 9  & 5.7 (± 0.0)     & 3.3           & 1 & 4          & 8.0 (±  0.6 )                          & 2.8 (±  1.1)  & \textbf{5.2}  & 1.0 (± 0.0)              \\ \hline
BM5  & 6  & \textbf{3} & 6  & 2.4 (± 0.0)     & \textbf{3.6}  & 1 & 7          & 5.0 (±  0.0 )                          & 1.9 (± 1.1)  & 3.1           & \textbf{1.0 (± 1.2)}     \\ \hline
BM6  & 8  & 3          & 8  & 3.0 (± 0.0)       & 2.0             & 1 & 4          & 6.7 (±  1.2 )                & 2.5 (± 0.5)  & \textbf{4.2}  & \textbf{88.5 (±7.5)}    \\ \hline
BM8 & 8 & 4 & 8 & 3.4 (± 0) & \textbf{4.6} & 1 & 6 & 8.0 (± 0) & 4.1 (± 1.1) & 3.9 & 1.0 (± 0.0) \\ \hline

\end{tabu}
}
\label{table:lime-vs-hfdt}
\end{table*}

\noindent \textbf{Explanations via \toolname (RQ3).}
\label{sec:rq3}
Our goal is to find out whether \toolname provides robust and succinct explanation compared to the baseline \textsc{LIME}~\cite{ribeiro2016should}. Given a set of inputs that characterize the significant discrimination (a high value of $K$), we investigate the performance of \toolname (explanation) to \textsc{LIME}. We note that \textsc{LIME} cannot provide the explanation out-of-the-box since its goal is to explain the outcome for one sample. To enable \textsc{LIME} to provide such explanations, we query it with $K$ samples and extract the common features from the top $3$ features in each explanation. To study the robustness of \textsc{LIME}, we perturb
the important features (one by one), measure the value of $k$-discrimination by querying the DNN model on the perturbed features, and report the difference between the initial and perturbed values.

We report the results of experiments in Table~\ref{table:lime-vs-hfdt}, with the columns \texttt{Init.K} - the max $k$-discrimination value from the search step, \texttt{Size} - the size of explanation in terms of \# of conjunction, \texttt{K} - the mean $k$-value before perturbation, \texttt{Pert.K} - the $k$-value after perturbation of the corresponding features for robustness, \texttt{Diff} - the difference between the initial and perturb explanations, and \texttt{Cov} - the input area volume of explanation for generality. 
Let us first study the overall trend in Table~\ref{table:lime-vs-hfdt}. Considering the Adult census (AC) dataset,
\texttt{Init.K} and \texttt{K} values vary between 17 to 20, \texttt{Size} value varies between 1 to 13,  \texttt{Diff} value varies between 2.3 to 17.2, and \texttt{Cov} varies between 1 to 2,590.

Looking into Table~\ref{table:lime-vs-hfdt}, \toolname has given an explanation both with smaller or equal size, but with more robust validation rules in terms of diff values, and with significantly higher coverage.
There are very few cases when \toolname requires a larger size of explanation, but in those cases, the explanation rules show strong performances in terms of coverage and validation diff values. 
Specifically, \toolname performs better in more than 78\% of the cases with respect to the robustness of explanations.
In all cases, \toolname covers a larger volume of input space than LIME.
\toolname provides a succinct explanation, whereas LIME's explanations may use all features, which may make it difficult to find the root causes. 
\noindent \textbf{Concrete Example of Explanation.}
\toolname also provides rich explanation through the hyper-rectangular split of the input space 
to show the significant discrimination (see all decision tree models in the appendix). The following plot shows a DT that explains the circumstances under
\begin{wrapfigure}{t}{0.27\textwidth}
\vspace{-1.0 em}    
\begin{flushright}
    \centering
    \captionsetup{font=tiny}
    \includegraphics[width=0.27\textwidth]{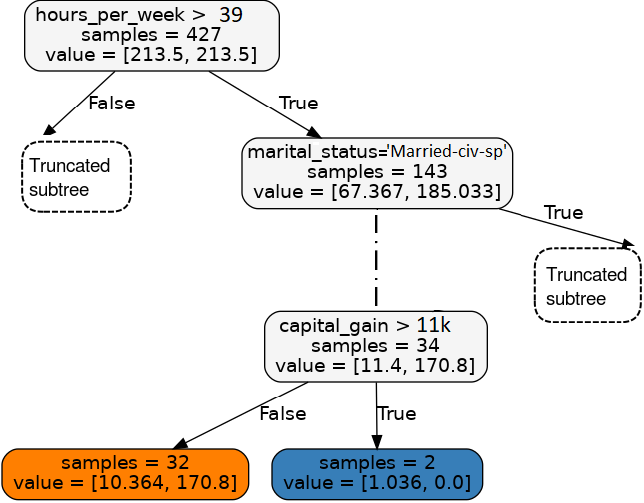}
    \label{fig:DT_overview}
\end{flushright}
\vspace{-1.0 em}    
\end{wrapfigure}
which the DNN model for the AC2 benchmark becomes significantly discriminatory. 
For example, the predicate \{hours\_per\_week $>$ 39 $\land$ marital\_status != [`Married-civ spouse']  $\land$ hours\_per\_week $<=$ 64 $\land$ workclass = [`Private', `Self-emp-not-inc', `Self-emp-inc', `Federal-gov'] $\land$ native\_country = [`United-States', `Cambodia', `England', `Puerto-Rico', `Canada', `Germany'] $\land$ capital\_gain $<=$ 11000\} shows a hyper-rectangular region with a significant arbitrariness (note: a subset of conditions are not shown in the DT due to the presentation). 

\begin{tcolorbox}[boxrule=1pt,left=1pt,right=1pt,top=1pt,bottom=1pt]
\textbf{Answer RQ3:} \toolname provides rich and succinct explanations that cover a larger sub-space and is more robust in explaining the root cause of significant arbitrariness. 
\end{tcolorbox}

\begin{table*}

    \centering
    \caption{RQ4: Original vs. Debiased model (with/without decision tree rules).}
    \label{tab:debiasing_perf} 
    \resizebox{0.95\textwidth}{!}{%
    \begin{tabu}{|l|l|l|l|l|l|l|l|l|l|}
        \hline
         DNN & Intervention & Acc (\%) & Iter($\times$ 1K) & $Max.K$ & $Avg.K$ & $\#ID$ & $Succ.rate$ (\%) & $\#ID.maxK.1$ \\ \hline
         \multirow{2}{4.5 em}{AC1} & Original & 81.81 & 3.6 ($\pm$ 1.1) & 20.0 ($\pm$ 0.0) & 9.3 ($\pm$ 1.1) & 1927.0 ($\pm$ 552.5) & 53.3 ($\pm$ 1.8) & 2.8 ($\pm$ 1.4) \\
         
         ~ & Original w DT & 81.81 & 3.7 ($\pm$ 1.2) & \textbf{18.1} ($\pm$ 0.3) & \textbf{9.1} ($\pm$ 1.0) & 2316.0 ($\pm$ 357.4) & 68.6 ($\pm$ 23.2) & 17.5 ($\pm$ 8.8) \\
         
          & Debias w/o DT & 81.22 & 3.6 ($\pm$ 0.0) & 20.0 ($\pm$ 0.0) & 13.2 ($\pm$ 0.1) & 1172.3 ($\pm$ 9.2) & 32.5 ($\pm$ 0.2) & 1.7 ($\pm$ 0.6) \\
          
         ~ & Debias w DT & 81.22 & 3.6 ($\pm$ 0.0) & 19.5 ($\pm$ 0.8) & 13.6 ($\pm$ 0.4) & \textbf{1112.7} ($\pm$ 605.7) & \textbf{30.9} ($\pm$ 16.8) & 4.6 ($\pm$ 5.0) \\
        \hline
        
         \multirow{2}{4.5 em}{AC2} & Original & 83.19 &  2.6 ($\pm$ 0.4) & 19.0 ($\pm$ 0.0) & 14.0 ($\pm$ 0.8) & 2245.0 ($\pm$ 318.8) & 87.7 ($\pm$ 0.9) & 3.8 ($\pm$ 2.2) \\
         ~ & Original w DT & 83.19  & 2.5 ($\pm$ 0.4) & \textbf{18.2} ($\pm$ 0.4) & 14.2 ($\pm$ 0.1) & 955.0 ($\pm$ 131.9) & 37.6 ($\pm$ 2.1) & \textbf{1.5} ($\pm$ 0.7) \\
         
          ~ & Debias w/o DT & 82.32 & 2.6 ($\pm$ 0.0) & 20.0 ($\pm$ 0.0) & \textbf{4.8} ($\pm$ 0.0) & 1244.0 ($\pm$ 0.0) & 48.5 ($\pm$ 0.0) & 3.0 ($\pm$ 0.0) \\
         ~ & Debias w DT & 82.32 & 2.2 ($\pm$ 0.5) & 19.5 ($\pm$ 0.5) & 5.2 ($\pm$ 0.5) & \textbf{836.0} ($\pm$ 304.1) & 37.8 ($\pm$ 10.1) & 4.2 ($\pm$ 1.7) \\

         \hline
         
         \multirow{2}{4.5 em}{AC3} & Original & 83.29 & 3.2 ($\pm$ 0.9) & 20.0 ($\pm$ 0.0) & 14.3 ($\pm$ 2.2) & 1794.1 ($\pm$ 380.5) & 58.0 ($\pm$ 6.3) & 166.8 ($\pm$ 69.8) \\
         ~ & Original w DT & 83.29  & 3.2 ($\pm$ 0.9) & \textbf{19.7} ($\pm$ 0.5) & \textbf{11.9} ($\pm$ 3.9) & \textbf{82.1} ($\pm$ 36.8) & \textbf{2.8} ($\pm$ 1.6) & \textbf{2.5} ($\pm$ 1.1) \\
         ~ & Debias w/o DT & 82.25 & 3.2 ($\pm$ 0.0) & 20.0 ($\pm$ 0.0) & 16.8 ($\pm$ 0.0) & 2230.0 ($\pm$ 0.0) & 70.5 ($\pm$ 0.0) & 490.0 ($\pm$ 0.0) \\
         
         ~ & Debias w DT & 82.25 & 3.2 ($\pm$ 0.0) & 20.0 ($\pm$ 0.0) & 16.8 ($\pm$ 0.4) & 1890.7 ($\pm$ 556.4) & 59.7 ($\pm$ 17.6) & 422.8 ($\pm$ 111.7) \\
         \hline
         
         \multirow{2}{4.5 em}{AC4} & Original & 82.74 & 0.1 ($\pm$ 0.0) & 20.0 ($\pm$ 0.0) & 17.1 ($\pm$ 0.4) & 124.8 ($\pm$ 7.3) & 90.5 ($\pm$ 4.6) & 13.2 ($\pm$ 4.9) \\
         ~ & Original w DT & 82.74 &  0.1 ($\pm$ 0.0) & 16.9 ($\pm$ 2.2) & 12.9 ($\pm$ 2.9) & 3.8 ($\pm$ 2.1) & 2.8 ($\pm$ 1.5) & 0.9 ($\pm$ 0.3) \\
         ~ & Debias w/o DT & 82.60 & 0.1 ($\pm$ 0.0) & \textbf{5.0} ($\pm$ 10.0) & \textbf{5.0} ($\pm$ 10.0) & \textbf{0.2} ($\pm$ 0.5) & \textbf{0.2} ($\pm$ 0.4) & \textbf{0.2} ($\pm$ 0.5) \\
         
         ~ & Debias w DT & 82.60 & 0.1 ($\pm$ 0.0) & 11.8 ($\pm$ 10.8) & 9.2 ($\pm$ 8.4) & 18.5 ($\pm$ 21.9) & 13.2 ($\pm$ 15.5) & 2.7 ($\pm$ 3.4) \\
         
         \hline
         
         \multirow{2}{4.5 em}{AC5} & Original & 83.36 & 1.6 ($\pm$ 0.2) & 18.2 ($\pm$ 0.5) & 13.0 ($\pm$ 0.1) & 1545.2 ($\pm$ 235.1) & 97.7 ($\pm$ 0.8) & 9.6 ($\pm$ 5.7) \\
         ~ & Original w DT & 83.36 & 1.6 ($\pm$ 0.2) & \textbf{17.5} ($\pm$ 0.7) & 13.5 ($\pm$ 0.1) & 866.5 ($\pm$ 141.3) & 55.1 ($\pm$ 1.4) & \textbf{2.4} ($\pm$ 2.0) \\
         ~ & Debias w/o DT & 82.66 & 1.3 ($\pm$ 0.4) & 18.5 ($\pm$ 2.1) & 11.1 ($\pm$ 0.9) & 696.0 ($\pm$ 79.2) & 57.3 ($\pm$ 13.9) & 15.5 ($\pm$ 17.7) \\
         
         ~ & Debias w DT & 82.66 & 1.4 ($\pm$ 0.3) & 19.5 ($\pm$ 1.7) & \textbf{10.7} ($\pm$ 0.8) & \textbf{684.8} ($\pm$ 79.6) & \textbf{49.2} ($\pm$ 7.3) & 15.2 ($\pm$ 14.2) \\
         \hline
         
         \multirow{2}{4.5 em}{AC6} & Original & 82.04 & 2.9 ($\pm$ 0.5) & 20.0 ($\pm$ 0.0) & 13.8 ($\pm$ 0.3) & 2623.2 ($\pm$ 499.7) & 91.5 ($\pm$ 1.5) & 108.4 ($\pm$ 25.7) \\
         ~ & Original w DT & 82.04 & 2.8 ($\pm$ 0.6) & \textbf{18.8} ($\pm$ 0.9) & 13.6 ($\pm$ 0.3) & \textbf{208.7} ($\pm$ 38.7) & \textbf{7.4} ($\pm$ 0.6) & \textbf{2.9} ($\pm$ 3.7) \\
         
         ~ & Debias w/o DT & 80.86 & 2.9 ($\pm$ 0.0) & 20.0 ($\pm$ 0.0) & 10.3 ($\pm$ 0.0) & 312.0 ($\pm$ 0.0) & 10.9 ($\pm$ 0.0) & 2.0 ($\pm$ 0.0) \\
         ~ & Debias w DT & 80.86 & 2.9 ($\pm$ 0.0) & 19.2 ($\pm$ 1.2) & \textbf{10.0} ($\pm$ 0.2) & 285.7 ($\pm$ 27.6) & 10.0 ($\pm$ 1.0) & 2.0 ($\pm$ 1.1) \\

         \hline
         
         \multirow{2}{4.5 em}{AC7} & Original & 82.96 & 4.3 ($\pm$ 0.9) & 16.5 ($\pm$ 0.5) & 14.3 ($\pm$ 0.4) & 17.9 ($\pm$ 6.7) & 0.4 ($\pm$ 0.1) & 2.6 ($\pm$ 1.6) \\
         ~ & Original w DT & 82.96 & 4.3 ($\pm$ 0.9) & 11.0 ($\pm$ 0.0) & 10.2 ($\pm$ 1.1) & 0.3 ($\pm$ 0.6) & 0.0 ($\pm$ 0.0) & 0.2 ($\pm$ 0.4) \\
         ~ & Debias w/o DT & 81.91 & 3.2 ($\pm$ 0.0) & \textbf{0.0} ($\pm$ 0.0) &\textbf{0.0} ($\pm$ 0.0) & \textbf{0.0} ($\pm$ 0.0) & \textbf{0.0} ($\pm$ 0.0) & \textbf{0.0} ($\pm$ 0.0) \\
         ~ & Debias w DT & 81.91 & 2.3 ($\pm$ 1.3) & 9.5 ($\pm$ 13.4) & 8.3 ($\pm$ 11.7) & 9.0 ($\pm$ 12.7) & 0.6 ($\pm$ 0.9) & 2.0 ($\pm$ 2.8) \\

         \hline

         \multirow{2}{4.5 em}{AC8} & Original & 82.45 & 3.8 ($\pm$ 1.5) & 16.9 ($\pm$ 0.3) & 12.3 ($\pm$ 0.2) & 2071.6 ($\pm$ 517.7) & 57.0 ($\pm$ 8.8) & 9.2 ($\pm$ 3.6) \\
          & Original w DT & 82.45 & 3.9 ($\pm$ 1.5) & \textbf{16.0} ($\pm$ 0.0) &  \textbf{11.7} ($\pm$ 0.6) &  \textbf{856.9} ($\pm$ 160.3) &  \textbf{24.1} ($\pm$ 7.6) &  \textbf{5.2} ($\pm$ 2.6) \\
         ~ & Debias w/o DT & 81.65 & 3.8 ($\pm$ 0.0) & 20.0 ($\pm$ 0.0) & 18.3 ($\pm$ 0.1) & 3088.5 ($\pm$ 20.5) & 81.3 ($\pm$ 0.6) & 1255.0 ($\pm$ 25.5) \\
         ~ & Debias w DT & 81.65 & 3.8 ($\pm$ 0.0) & 20.0 ($\pm$ 0.0) & 18.4 ($\pm$ 0.1) & 3135.0 ($\pm$ 1.4) & 82.4 ($\pm$ 0.1) & 1338.5 ($\pm$ 113.8) \\

         \hline
         
         \multirow{2}{4.5 em}{AC9} & Original & 82.07 & 3.1 ($\pm$ 0.4) & 19.0 ($\pm$ 0.0) & 13.7 ($\pm$ 0.7) & 2430.4 ($\pm$ 323.3) & 77.6 ($\pm$ 1.4) & 51.0 ($\pm$ 8.9) \\
         ~ & Original w DT &  82.07 & 3.2 ($\pm$ 0.5) & \textbf{18.0} ($\pm$ 0.0) & \textbf{13.5} ($\pm$ 1.2) &  \textbf{102.6} ($\pm$ 20.3) &  \textbf{3.3} ($\pm$ 0.8) &  \textbf{7.6} ($\pm$ 2.0) \\
         ~ & Debias w/o DT & 80.96 & 3.1 ($\pm$ 0.0) & 19.0 ($\pm$ 0.0) & 16.4 ($\pm$ 0.0) & 2766.0 ($\pm$ 0.0) & 88.2 ($\pm$ 0.0) & 25.0 ($\pm$ 0.0) \\
         ~ & Debias w DT & 80.96 & 3.1 ($\pm$ 0.0) & 18.2 ($\pm$ 1.5) & 16.0 ($\pm$ 0.5) & 2051.8 ($\pm$ 1367.1) & 65.4 ($\pm$ 43.6) & 14.2 ($\pm$ 9.9) \\

         \hline
         
         \multirow{2}{4.5 em}{AC10} & Original & 81.68 & 3.5 ($\pm$ 0.9) & 16.9 ($\pm$ 0.3) & \textbf{11.3} ($\pm$ 1.6) & 1838.2 ($\pm$ 544.9) & 51.6 ($\pm$ 2.1) & 5.8 ($\pm$ 11.5) \\
         ~ & Original w DT & 81.68 & 3.6 ($\pm$ 1.0) & \textbf{16.0} ($\pm$ 0.0) & 13.0 ($\pm$ 0.3) &  \textbf{804.2} ($\pm$ 279.5) &  \textbf{22.0} ($\pm$ 2.1) &  \textbf{8.5} ($\pm$ 2.8) \\
         ~ & Debias w/o DT & 81.28 & 3.5 ($\pm$ 0.0) & 19.0 ($\pm$ 0.0) & 12.5 ($\pm$ 0.0) & 2101.0 ($\pm$ 0.0) & 59.4 ($\pm$ 0.0) & 104.0 ($\pm$ 0.0) \\
         ~ & Debias w DT & 81.28 & 3.5 ($\pm$ 0.0) & 19.0 ($\pm$ 0.0) & 12.5 ($\pm$ 0.1) & 2145.2 ($\pm$ 106.5) & 60.6 ($\pm$ 3.0) & 99.2 ($\pm$ 9.2) \\

         \hline
         
         \multirow{2}{4.5 em}{AC11} & Original & 82.19 & 1.3 ($\pm$ 0.3) & 20.0 ($\pm$ 0.0) & 15.0 ($\pm$ 0.4) & 987.3 ($\pm$ 274.9) & 73.8 ($\pm$ 7.1) & 29.9 ($\pm$ 16.2) \\
         ~ & Original w DT & 81.19 & 1.3 ($\pm$ 0.3) & 19.6 ($\pm$ 0.5) & 14.0 ($\pm$ 0.8) &  \textbf{435.1} ($\pm$ 123.2) &  \textbf{32.7} ($\pm$ 2.6) & 9.6 ($\pm$ 11.9) \\
         
         ~ & Debias w/o DT & 80.85 & 1.3 ($\pm$ 0.0) & 20.0 ($\pm$ 0.0) & 10.2 ($\pm$ 0.0) & 767.0 ($\pm$ 0.0) & 57.5 ($\pm$ 0.0) & 6.0 ($\pm$ 0.0) \\
         
         ~ & Debias w DT & 80.85 & 1.3 ($\pm$ 0.0) & \textbf{17.7} ($\pm$ 4.9) &  \textbf{9.2} ($\pm$ 1.5) & 613.7 ($\pm$ 521.2) & 46.0 ($\pm$ 39.1) &  \textbf{3.3} ($\pm$ 2.5) \\

         \hline

         \multirow{2}{4.5 em}{AC12} & Original & 82.16 & 2.4 ($\pm$ 0.9) & 20.0 ($\pm$ 0.0) & 15.7 ($\pm$ 1.7) & 1661.8 ($\pm$ 593.3) & 70.8 ($\pm$ 1.6) & 60.6 ($\pm$ 25.6) \\
         ~ & Original w DT & 82.16 & 2.4 ($\pm$ 0.9) & 19.2 ($\pm$ 0.6) & 17.7 ($\pm$ 0.6) &  \textbf{6.8} ($\pm$ 2.7) &  \textbf{0.3} ($\pm$ 0.1) & 1.8 ($\pm$ 1.5) \\
         ~ & Debias w/o DT & 81.66 &  2.4 ($\pm$ 0.0) & 15.0 ($\pm$ 0.0) &  \textbf{9.2} ($\pm$ 0.0) & 1483.0 ($\pm$ 0.0) & 62.8 ($\pm$ 0.0) & 2.0 ($\pm$ 0.0) \\
         ~ & Debias w DT & 81.66 & 2.4 ($\pm$ 0.0) & \textbf{14.2} ($\pm$ 0.5) &  \textbf{9.2} ($\pm$ 0.1) &  1111.5 ($\pm$ 718.5) & 47.1 ($\pm$ 30.4) & 9.8 ($\pm$ 9.8) \\
         \hline


         \multirow{2}{4.5 em}{BM1} & Original & 90.63 &  9.5 ($\pm$ 1.5) & 9.0 ($\pm$ 0.0) & 5.2 ($\pm$ 0.6) & 5357.6 ($\pm$ 391.0) & 57.2 ($\pm$ 6.2) & 28.3 ($\pm$ 7.9) \\
          ~ & Original w DT & 90.63 & 10.0 ($\pm$ 1.5) & \textbf{5.8} ($\pm$ 0.4) &  \textbf{4.7} ($\pm$ 0.4) &  \textbf{21.8} ($\pm$ 4.8) &  \textbf{0.2} ($\pm$ 0.1) &  \textbf{2.6} ($\pm$ 2.9) \\
         ~ & Debias w/o DT & 90.11 & 5.9 ($\pm$ 0.0) & 9.0 ($\pm$ 0.0) & 6.1 ($\pm$ 0.0) & 335.5 ($\pm$ 57.3) & 5.6 ($\pm$ 0.9) & 26.0 ($\pm$ 7.1) \\
         ~ & Debias w DT & 90.11 & 4.1 ($\pm$ 0.7) & 8.5 ($\pm$ 0.6) & 5.8 ($\pm$ 0.4) & 136.8 ($\pm$ 116.6) & 3.5 ($\pm$ 3.0) & 10.8 ($\pm$ 6.6) \\
         \hline

         \multirow{2}{4.5 em}{BM2} & Original & 90.27 &  11.3 ($\pm$ 0.0) & 9.0 ($\pm$ 0.0) & 4.6 ($\pm$ 1.1) & 730.0 ($\pm$ 195.9) & 6.5 ($\pm$ 1.7) & 18.4 ($\pm$ 5.8) \\   
        
         ~ & Original w DT & 90.63 & 11.3 ($\pm$ 0.0) & 8.4 ($\pm$ 0.5) & 5.1 ($\pm$ 0.8) &  \textbf{727.2} ($\pm$ 760.7) &  \textbf{6.4} ($\pm$ 6.7) & 5.5 ($\pm$ 8.7) \\
         ~ & Debias w/o DT & 89.86 & 11.3 ($\pm$ 0.0) & 9.0 ($\pm$ 0.0) & 6.0 ($\pm$ 0.0) & 2168.5 ($\pm$ 34.6) & 19.2 ($\pm$ 0.3) & 53.5 ($\pm$ 0.7) \\
         ~ & Debias w DT & 89.86 & 7.2 ($\pm$ 5.8) & \textbf{4.5} ($\pm$ 6.4) & \textbf{2.9} ($\pm$ 4.2) & 750.5 ($\pm$ 1061.4) & 6.6 ($\pm$ 9.4) & 9.0 ($\pm$ 12.7) \\         
         \hline
         
         \multirow{2}{4.5 em}{BM3} & Original & 90.35  & 11.3 ($\pm$ 0.1) & 9.0 ($\pm$ 0.0) & 6.5 ($\pm$ 0.9) & 2988.4 ($\pm$ 311.6) & 26.5 ($\pm$ 2.9) & 422.8 ($\pm$ 128.3) \\ 
         ~ & Original w DT &  90.35 &  11.3 ($\pm$ 0.1) & \textbf{7.2} ($\pm$ 0.4) &  \textbf{5.3} ($\pm$ 0.9) &  \textbf{12.9} ($\pm$ 9.9) &  \textbf{0.1} ($\pm$ 0.1) &  \textbf{3.5} ($\pm$ 2.5) \\
         
         ~ & Debias w/o DT & 90.17 & 6.7 ($\pm$ 1.0) & 9.0 ($\pm$ 0.0) & 7.8 ($\pm$ 0.0) & 1296.5 ($\pm$ 207.2) & 19.2 ($\pm$ 0.2) & 386.0 ($\pm$ 91.9) \\
 
         ~ & Debias w DT & 90.17  &  5.4 ($\pm$ 1.3) & 9.0 ($\pm$ 0.0) & 7.5 ($\pm$ 0.0) & 326.0 ($\pm$ 387.2) & 5.5 ($\pm$ 6.4) & 76.2 ($\pm$ 90.3) \\

         \hline
         
         \multirow{2}{4.5 em}{BM5} & Original & 90.26 & 10.2 ($\pm$ 1.7) & 6.0 ($\pm$ 0.0) & 3.7 ($\pm$ 0.5) & 1608.0 ($\pm$ 248.6) & 16.0 ($\pm$ 2.4) & 25.1 ($\pm$ 22.4) \\    
         ~ & Original w DT & 90.26 & 11.3 ($\pm$ 0.0) & \textbf{5.0} ($\pm$ 0.0) & 4.1 ($\pm$ 0.0) & 520.0 ($\pm$ 17.9) & 4.6 ($\pm$ 0.2) & 150.0 ($\pm$ 19.7) \\
         ~ & Debias w/o DT & 89.81 & 10.2 ($\pm$ 0.0) & 9.0 ($\pm$ 0.0) & 4.7 ($\pm$ 0.0) & 264.0 ($\pm$ 26.9) & 2.6 ($\pm$ 0.3) & 2.0 ($\pm$ 0.0) \\
         ~ & Debias w DT & 89.81 & 10.2 ($\pm$ 0.0) & 6.5 ($\pm$ 4.4) & \textbf{3.5} ($\pm$ 2.3) & \textbf{182.2} ($\pm$ 121.6) & \textbf{1.8} ($\pm$ 1.2) & 3.8 ($\pm$ 4.9) \\
         \hline
         
         \multirow{2}{4.5 em}{BM6} & Original & 89.89 & 11.3 ($\pm$ 0.0) & 8.0 ($\pm$ 0.0) & \textbf{3.4} ($\pm$ 0.0) & 2315.0 ($\pm$ 328.3) & 20.5 ($\pm$ 2.9) & \textbf{21.5} ($\pm$ 13.1) \\          
         ~ & Original w DT & 89.89 & 11.3 ($\pm$ 0.0) & \textbf{6.0} ($\pm$ 0.0) & 2.6 ($\pm$ 0.0) & \textbf{608.2} ($\pm$ 38.1) & \textbf{5.4} ($\pm$ 0.3) & 26.8 ($\pm$ 4.7) \\
         ~ & Debias w/o DT & 89.03 & 11.3 ($\pm$ 0.0) & 9.0 ($\pm$ 0.0) & 6.1 ($\pm$ 0.0) & 2902.5 ($\pm$ 16.3) & 25.7 ($\pm$ 0.1) & 56.0 ($\pm$ 2.8) \\
         ~ & Debias w DT & 89.03 & 11.3 ($\pm$ 0.0) & 8.8 ($\pm$ 0.5) & 6.4 ($\pm$ 0.4) & 1963.2 ($\pm$ 1346.0) & 17.4 ($\pm$ 11.9) & 32.8 ($\pm$ 21.9) \\         
         \hline
         
         \multirow{2}{4.5 em}{BM8} & Original & 90.07 & 0.3 ($\pm$ 4.7) & 8.0 ($\pm$ 0.0) & 4.8 ($\pm$ 0.1) & 41.8 ($\pm$ 4.4) & 16.5 ($\pm$ 3.5) & 4.2 ($\pm$ 1.5) \\    
         ~ & Original w DT & 90.07 & 0.3 ($\pm$ 0.1) & 5.0 ($\pm$ 0.0) & 3.9 ($\pm$ 0.2) & 26.2 ($\pm$ 4.0) & 10.2 ($\pm$ 1.6) & 4.5 ($\pm$ 1.7) \\
         ~ & Debias w/o DT & 90.22 & 3.8 ($\pm$ 3.0) & 3.0 ($\pm$ 4.2) & 3.0 ($\pm$ 4.2) & 0.5 ($\pm$ 0.7) & 0.0 ($\pm$ 0.0) & 0.5 ($\pm$ 0.7) \\
         ~ & Debias w DT & 90.22 & 5.6 ($\pm$ 2.0) & \textbf{0.0} ($\pm$ 0.0) &  \textbf{0.0} ($\pm$ 0.1) &  \textbf{0.1} ($\pm$ 0.1) &  \textbf{0.0} ($\pm$ 0.0) &  \textbf{0.2} ($\pm$ 0.2) \\
         \hline
         
    \end{tabu}
    }
\end{table*}

\noindent \textbf{Debiasing unfairness via \toolname (RQ4).}
\label{sec:rq4}
We perform two interventions to mitigate unfairness in the original models: 1) we add decision tree rules as guardrails to refute queries that can lead to arbitrary outcomes, and 2) we retrain the original DNN models over curated discriminatory instances via data augmentation, similar to prior works~\cite{zhaoaft2024ase,fanExpGAICSE22,zhang2020white,udeshi2018automated} to obtain debiased models.
Table~\ref{tab:debiasing_perf} shows the comparison between the original and debiased models with and without decision tree rules. Overall, the debiased models led to at most 2\% reduction in the accuracy. 
We apply simulated annealing (SA) to search for unfairness in the mitigated models, similar to the original ones. 
We find that adding the decision tree rules to the original models outperform other techniques in reducing the maximum $k$-discrimination in 67\% of cases. The debiased models with and without DT achieved better results in 22\% and 11\% of cases, respectively. Similarly, adding DT rules to the original models reduce the success rates of finding individual discrimination cases in 67\% of cases where the debiased models with DT achieved better results in 28\% of cases. 
When considering the average $k$-discrimination of ID samples, both original and debiased models with DT tie by reducing it in 39\% of cases.

One interesting and unexpected outcome is that while  the debiased models reduce the $Succ.rate$ and $\#ID$, they often increase $Max.K$ value. This shows that reducing $k$-discrimination with simple retraining strategies does not work and requires careful retraining and novel strategies. While adding decision tree rules help, we believe that retraining introduces new fairness vulnerabilities that require further iteration of \toolname to infer new discriminatory rules.  We believe that more in-depth future research is necessary for debiasing k-discrimination bugs in the models.

\begin{tcolorbox}[boxrule=1pt,left=1pt,right=1pt,top=1pt,bottom=1pt]
\textbf{Answer RQ4:} Applying decision rules as guardrails for denying output in more sensitive cases with the original model (67\%) and retrained models (22\%) reduces $k$-discrimination metrics in 89\% of cases.
\end{tcolorbox}

\section{Discussion}
\label{sec:discussion}

\noindent \textit{Limitation}.
In this work for generating counterfactuals, we perturb for all possible combinations of the protected attribute which might lead to some unrealistic or imaginary counterfactual instances. 
We use some rule-of-thumb relationships to mitigate this issue (e.g., a married individual with a female gender cannot be husband for the relationship attribute), but Conditional GANs and Variational Auto Encoders can be employed to improve the realism of samples~\cite{xiao2023latent}.

Our current technique also does not handle intersectional fairness~\cite{10.1145/3540250.3549103,10.1145/3597503.3639083,pmlr-v142-ghosh21a}, which reveals unfairness in the combination of multiple protected attributes. To overcome this limitation, our proposed approach can be repeated multiple times (one per each combination) to certify fairness or find the maximum unfairness. 


\vspace{0.5em}
\noindent \textit{Threat to Validity.} To ensure the validity of our experiments and the reproducibility and valid conclusion, we follow established rules and guidelines and take the average of the repeated experiments to validate our claims. To ensure that our results are generalizable and address external validity, and apply to multiple datasets, we experiment on $20$ DNN models taken from the literature of fairness testing and the real world in Kaggle and use $2$ different datasets. Our certification is limited to a given fairness notion, bounded to a time-out, and sensitive to seed selection. Hence, we may not guarantee fairness in general. Decision tree algorithms have the limitation of hyper-rectangular partitioning and may not show the causal relationships between input features and discriminatory instances.

\section{Related Work}
\label{sec:related}

\noindent \textbf{Verifying Fairness Properties.}
Multiple prior works used formal techniques to certify fairness in the ML models~\cite{john2020verifying,albarghouthi2017fairsquare,khedr2023certifair,10.1109/ICSE48619.2023.00134,kim2024fairquantcertifyingquantifyingfairness,monjezi2024causal}.
\textsc{Fairify}~\cite{10.1109/ICSE48619.2023.00134} addressed the fairness verification problem of neural networks for individual fairness. They formulate pre-trained DNNs via Satisfiability modulo theories (SMT), and either certify the DNN for individual fairness or find a counterexample that is an instance of individual discrimination. However, these approaches cannot distinguish between counterexamples, which are critical for prioritizing counterexamples and explaining patterns in fairness bugs. 

\noindent \textbf{Testing Fairness Properties.}
Multiple research works~\cite{angell2018themis,agarwal2018automated,10.1145/3338906.3338937,10.1145/3510003.3510137,monjezi2025fairness}
consider testing the individual discrimination in non-neuron ML models. 
\textsc{Themis}~\cite{angell2018themis}, \textsc{AEQUITAS}~\cite{udeshi2018automated}, \textsc{ADF}~\cite{zhang2020white}, \textsc{AFT}~\cite{zhaoaft2024ase}, \textsc{EXPGA}~\cite{fanExpGAICSE22}, \textsc{NeuronFair}~\cite{9793943}, and EIDIG~\cite{10.1145/3460319.3464820} used causal fairness definition (different variants of $2$-fairness notions) that may not prioritize test cases and quantify different risks of harm.
DICE~\cite{Monjezi2023InformationTheoreticTA} employs an information theory-based method to quantify individual discrimination. However, DICE cannot guarantee the absence of unfairness. We use formal techniques to certify fairness and explain the root cause of bugs.  

\noindent \textbf{eXplainable AI.} 
\textsc{Parfait-ML}~\cite{tizpaz2022fairness} used decision trees to explain what hyperparameter configuration of ML libraries can lead to inferring unfair ML models. 
Mothilal et al. \cite{10.1145/3351095.3372850} provided diverse counterfactual explanations for a given decision subject that enables them to flip an ML decision outcome. Watcher et al.~\cite{wachter2018counterfactualexplanationsopeningblack} focused on understanding the decision flip by the feature-perturbed version of the same individual. 
\textsc{LORE}~\cite{guidotti2018localrulebasedexplanationsblack} used a decision tree to approximate the non-linear models, whereas \textsc{Anchors}~\cite{10.5555/3504035.3504222} used model-agnostic explanations based on if-then rules.
\toolname is geared towards DNN software, and it goes beyond i) explanation of decision for one subject and ii) prevalent differential analysis.  

\section{Conclusion}
\label{sec:conclusion}
This paper presented a hybrid framework for fairness analysis of neural networks by combining testing and verification techniques. We introduced a quantitative generalization of individual discrimination and proposed a method to explain the conditions under which DNN models exhibit significant clustered discrimination. Our approach supports both the detection and mitigation of such fairness violations. An important direction for future work is to systematically assess the risk posed by automated decision-support systems when a small number of marginalized groups receive unfavorable outcomes, even as the system appears fair to the majority.

\noindent \textbf{Acknowledgment.} 
The authors thank ASE
reviewers for their time and invaluable feedback to improve this work. 
This project has been supported by NSF under Grant No. CNS-2230060, CNS-2527657, CNS-2230061, and CCF-2317207.

\bibliographystyle{ieeetr}
\bibliography{references}


\end{document}